\documentclass[]{jfm}
\usepackage{pdflscape} 
\usepackage{array}
\usepackage{multirow}
\usepackage{graphicx}
\usepackage{rotating}
\usepackage{newtxtext}
\usepackage{newtxmath}
\usepackage{natbib}
\usepackage{hyperref}
\usepackage{adjustbox}   
\usepackage{lscape}      
\hypersetup{
    colorlinks = true,
    urlcolor   = blue,
    citecolor  = black,
}
\usepackage{svg}

\newcommand{\RomanNumeralCaps}[1]
\linenumbers

\title{Gravity current fronts advancing along a heated wall}

\author{Stefano Lanzini\aff{1}
  \corresp{\email{stefano.lanzini@ec-lyon.fr}},
  Mathieu Creyssels \aff{1},
  Massimo Marro \aff{1},
  Samuel Vaux \aff{2},
 \and Pietro Salizzoni\aff{1,3}}

\affiliation{\aff{1} Ecole Centrale de Lyon, CNRS, Universite Claude Bernard Lyon 1, INSA Lyon,
LMFA, UMR5509, 69130 Ecully, France
\aff{2}Autorité de Sûreté Nucléaire et de Radioprotection (ASNR),
PSN-RES, SA2I, LIE, Cadarache, 13115 Saint-Paul-lez-Durance,
France
\aff{3}Department of Environmental, Land and Infrastructure Engineering (DIATI), Politecnico di Torino, Corso Duca degli
Abruzzi 24, 10129 Turin, Italy}

\begin{document}
\maketitle

\begin{abstract}

The propagation of sustained Boussinesq gravity-current fronts along a heated wall is examined through laboratory experiments. The gravity currents are generated by continuously supplying a mixture of air and carbon dioxide at the inlet of a rectangular channel, while uniform wall heating is provided by resistive fabrics. In addition to the Froude number, defined either at the source ($Fr_s$) or at the current front ($Fr$), we show that the front velocity is governed by $\Lambda_s$, the ratio of the two buoyancy fluxes per unit area driving the flow: that generated by wall heating and that supplied at the inlet.
As wall heating increases (i.e., as $\Lambda_s$ increases), the current front propagates more slowly. This slowdown results from the direct heating of the current, which reduces the buoyancy of its head, and from the interaction between the front and the thermal plumes generated by natural convection downstream. In the most strongly heated experiments, the front eventually stops at a distance from the source that is inversely proportional to $\Lambda_s$. In these cases, the buoyancy of the head changes sign, and the head evolves into a positively buoyant plume whose vertical extent continues to increase. The experimental results are compared with predictions from a lumped analytical model describing the propagation and eventual arrest of the current.

\end{abstract}
\section{Introduction}\label{sec:introduction}
Gravity currents are predominantly horizontal flows driven by buoyancy differences between two fluids. Experimental and theoretical research on this phenomenon has been extensive, as it is ubiquitous in both natural and man-made environments \citep{Simpson1982}.  Gravity currents can be divided into two categories: those produced by the release of a finite mass of buoyant fluid into the ambient, usually referred to as lock-release currents, and those generated by a continuous supply of fluid, usually referred to as sustained currents. 
 The properties of both lock-release and sustained currents have been experimentally and numerically investigated under a wide range of ambient and wall conditions. Many studies have focused on the front propagation velocity and on the properties of the gravity current head.
 Beyond the widely studied case of a quiescent ambient fluid with homogeneous density (see the book of \cite{ungarish2020}), several laboratory studies have examined scenarios of current fronts advancing in a stably stratified \citep{MAXWORTHY2002, Longo2016,Chiapponi2018, laForgia} or rotating flow \citep{GriffithsHopfinger1983,Griffiths1986,Cenedese2004,CenedeseAdduce2008}, or in an ambient fluid with non-zero turbulence intensity \citep{LindenSimpson1986}.
Concerning the wall conditions, and focusing on flows advancing along the bottom wall of rectangular channels, the effects of free-slip \citep{Britter_Simpson_1978} and no-slip \citep{Simpson_Britter_1979} boundary conditions, roughness elements of different sizes and shapes (e.g., \cite{Nogueira2013,wilson2017,Jiang2018,Maggi2022,maggi2025}), and channel slope (e.g., \cite{Britter_Linden_1980,Negretti2017,Martin2020,DEFALCO2021}) on the properties of the gravity-current head have been investigated.


Despite the wide range of configurations investigated, the case of a non-adiabatic wall has so far received much less attention, despite its relevance to a wide range of real-world gravity currents.
 For example, the propagation of atmospheric gravity currents, such as sea-breeze fronts and thunderstorm outflows, is strongly influenced by sensible heat exchange with the Earth's surface. Field measurements of \cite{Simpson1977} show a deceleration of the sea-breeze front advancing velocity at noon and an acceleration in the late afternoon, as solar heating decreases. Other experimental \citep{CenedeseMonti2003} and numerical models (e.g., \cite{Yoshikado1992,Yoshikado1994,Thompsonetal,DiBernardino2022,Chen2025}) have shown that the propagation and structure of sea breeze fronts can be significantly altered when they interact with the urban heat island produced by cities. In this framework, \cite{Yoshikado1992} and \cite{CenedeseMonti2003} found that one of the relevant parameters that controls the head dynamics is the ratio between the horizontal speed of the sea breeze front and the convective velocity scale induced by the heating. 

Other examples arise at smaller spatial scales, such as the accidental release of light or dense gases. Hot smoke generated by tunnel fires forms a buoyant current that propagates beneath the tunnel ceiling and is progressively cooled by heat transfer to the relatively cold walls. This cooling can substantially alter smoke dispersion \citep{Salizzoni2018}, weakening the density stratification of the upper smoke layer and promoting its mixing throughout the tunnel cross-section. Conversely, accidental releases of cold industrial gases \citep{Britter1989, HANNA2012, FOX2022, Vidali2025} generate dense gravity currents that are heated by the ground, which is typically at ambient temperature. In all these cases, heat exchange with the boundary reduces the vertical density gradient within the current.

To take a step forward in understanding how the propagation of gravity current fronts can be influenced by a non-adiabatic boundary, this work examines the case of a sustained current propagating along a homogeneously heated wall.  In our setup, the current is generated by a mixture of air and carbon dioxide continuously supplied at the inlet of a channel featuring a homogeneously heated horizontal bottom wall. Both the wall-heating intensity and the inlet condition are systematically varied. 
This study aims to characterise the properties of the current by identifying the key non-dimensional parameters that govern its dynamics and evaluating their effects on the flow. Particular attention is devoted to the scaling laws describing the position of the front,  i.e., the current most forward point, as a function of time.
The understanding of the key physical processes governing the flow further enables the development of a lumped analytical model to predict the advance of the front in time, which is then compared with the experimental results.

The manuscript is organised as follows. In \S \ref{sec:dimensional}, the relevant non-dimensional governing parameters are identified. The experimental setup and the tested conditions are presented in \S\ref{sec:exp}. Flow visualizations and quantitative results are shown in \S \ref{sec:results}, and the derivation of the model is presented in \S \ref{sec:ch3model}.
The conclusions are discussed in \S \ref{sec:conclusion}.
\section{Governing parameters}
\label{sec:dimensional}

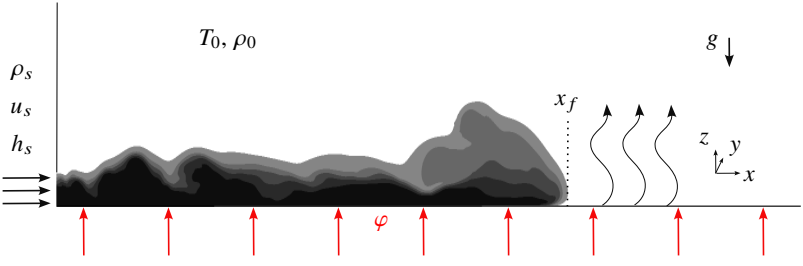
\begin{figure}

\centering
  \unitlength = 1.cm
\begin{picture}(12.5,3.5)
\put(0,0){\includegraphics[height=3.5cm]{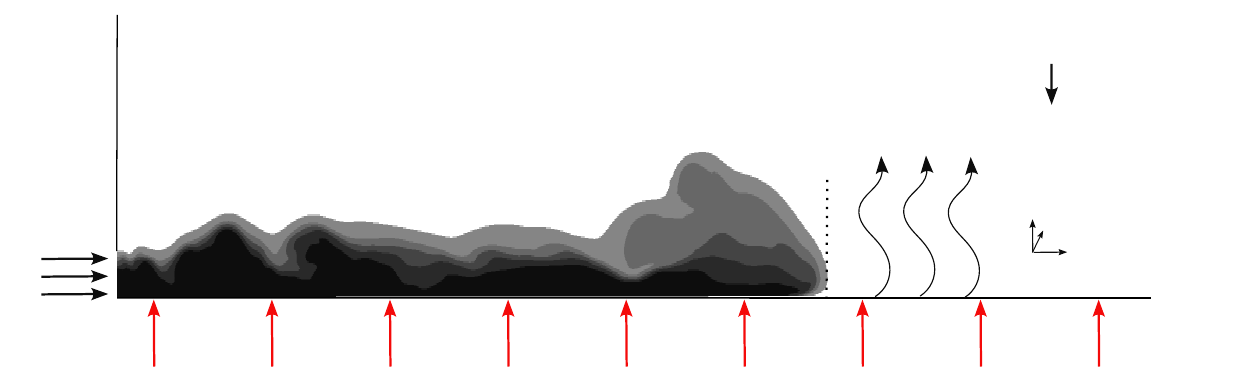}}
\put(0.5,2.4){$\rho_s$}
\put(0.5,1.9){$u_s$}
\put(0.5,1.4){$h_s$}
\put(3,2.8){$T_0$,  $\rho_0$}
\put(7.7,2){$x_f$}
\put(9.7,2.8){$g$}
\put(9.6,1.5){$z$}
\put(10.2,1){$x$}
\put(10,1.4){$y$}
\put(5.3,0.4){\textcolor{red}{$\varphi$}}

\end{picture}
   
    \caption{Sketch of a sustained gravity current advancing on a heating wall.}
    \label{fig:sketch}
\end{figure}

The first step of our study is a dimensional analysis to identify the relevant governing parameters of the system sketched in figure \ref{fig:sketch}, representing a sustained gravity current advancing on a heated horizontal wall. 
The current is generated by a constant release with horizontal velocity $u_s$ and density $\rho_s$,  from an inlet of height $h_s$ into an ambient that, far from the heating surface, has a density $\rho_0$ and temperature $T_0$.  For the sake of simplicity, possible differences between the momentum ($\nu$) and thermal ($\alpha$) diffusivities, and specific heats ($c_p$) of the two fluids (i.e., the ambient one and the dense release) are neglected \citep{lanzini2026}. This approximation is justified by the similar values of diffusivities of air and CO$_2$   and by the relatively small CO$_2$ volume fraction in the injected mixture (maximum 10\% in this work (\S \ref{sec:exp})).  
Prior to release, the bottom wall is uniformly heated until a statistically stationary state is attained, yielding a spatially uniform convective heat flux per unit area, $\varphi$, to the overlying air (\S \ref{sec:exp}). The dense current is generated only once this condition is established. Considering low-Mach number flow and ideal-gas behaviour for both the ambient and the current, after the mixture injection, the front position $x_f$ can be expressed as a function of the following parameters:
\begin{equation}\label{eq:dimensional}
    x_f = f\left(t, h_s, u_s,\rho_s,g, \rho_0,   T_0, \varphi,\nu,D_m, \alpha \right),
\end{equation}
where $t$ is the time after the injection, $g$ is the gravitational acceleration, and $D_m$ is the molecular diffusivity of carbon dioxide in air. The horizontal buoyancy flux per unit area at the source is defined as $f_s=u_sg(\rho_s-\rho_0)/\rho_s=u_sb_s$, and the source buoyancy flux per unit span is $F_s=f_s h_s$. The vertical buoyancy flux per unit area induced by the heating $\varphi$ is $ f_{\varphi} = g\varphi/\rho_0T_0c_p$ \citep{Linden1999}. Based on these definitions and on relation (\ref{eq:dimensional}), which involves 11 dimensional independent quantities with 4 independent dimensions (time, mass, length, temperature), the non-dimensional front position $x^*_f=x_f/h_s$ can be expressed as a function of the following 7 non-dimensional governing parameters \citep{Barenblatt_1996}:
\begin{equation}
   x^*_f = f\left(t^*,\frac{\rho_s}{\rho_0},Re_s, Fr_s, Sc, Pr, \Lambda_s \right),\label{eq:buckingam}
\end{equation}
where $t^*=t F_s^{1/3}/h_s$ is the non-dimensional time  \citep{SherWoods2017, Harrouk_Mehaddi_Arcen_Dossmann_2025}, $Re_s=u_sh_s/\nu$ is the source Reynolds number, $Fr_s=u_s /\sqrt{g(\rho_s-\rho_0)h_s/\rho_0}$ is the source Froude number,  $Sc=\nu/D_m$ and $Pr=\nu/\alpha$ are the Schmidt and Prandtl numbers, respectively, and $\Lambda_s=f_{\varphi}/f_s$.
This latter parameter quantifies the heating intensity and is defined as the ratio between the two buoyancy fluxes per unit area that characterise the problem (i.e., the vertical flux driven by heating, $f_{\varphi}$, and the horizontal flux that generates the current, $f_s$). 
Hereafter, the Prandtl and Schmidt numbers are kept fixed, and their influence on the front properties is not investigated.
A major advantage of using a heated air--CO$_2$ mixture is that double-diffusive instabilities, such as salt fingering, are not expected to develop. Although temperature and CO$_2$ concentration produce opposing contributions to the vertical density gradient, their molecular diffusivities are similar ($D_m = 1.6 \times 10^{-5}$ m$^2$ s$^{-1}$ and $\alpha = 2.0 \times 10^{-5}$ m$^2$ s$^{-1}$ at 20 $^\circ$C and atmospheric pressure, respectively \citep{Pritchard1982,Massman1998,lanzini2026}), whereas double-diffusive instabilities require substantially different scalar diffusivities \citep{Turner1974}. Consequently, such instabilities are not expected to influence the flow dynamics, making the present configuration a suitable analogue for gravity currents driven by a single scalar, namely temperature.

The analysis is limited to Boussinesq currents, with $\rho_s/\rho_0<1.052$. Preliminary adiabatic experiments were performed to determine the Reynolds number threshold above which the front velocity becomes independent of $Re_s$ (see \S \ref{subsec:frontposition}). Based on these results, all subsequent experiments were conducted at $Re_s \approx 1600$, well above the identified threshold. These values are consistent with previous investigations of sustained gravity currents, which found that the front velocity becomes effectively independent of the Reynolds number for $Re_s \gtrsim 1000$ \citep{Simpson1997, Hoggetal2016}. In view of the independence from the Reynolds number and density ratio, and with $Sc$ and $Pr$ held fixed, Equation (\ref{eq:buckingam}) reduces to
\begin{equation}\label{eq:goal}
x^*_f = f\left(t^*,Fr_s,\Lambda_s\right).
\end{equation}
The explicit dependence on time, $t^*$, can be recast in an implicit form by considering the time-dependent head Froude number (rather than the source value, $Fr_s$), defined as
\begin{equation}\label{eq:Fr}
    Fr=\frac{\text d x_f/ \text d t}{(b_h h_h)^{1/2}},
\end{equation}
where $b_hh_h$ is the vertical integral buoyancy immediately behind the front, evaluated at the location where the head reaches its maximum height, $h_h$ \citep{SherWoods2017}. For adiabatic, turbulent, Boussinesq, sustained gravity currents flowing on horizontal walls, \citet{SherWoods2017} showed that, after the front generation, the head Froude number becomes independent of both the source conditions and time, with an approximately constant value of $Fr=Fr_0\approx1.1\pm0.05$. 
In the presence of wall heating, however, $Fr$ is expected to be a function of both $t^*$ and $\Lambda_s$. Therefore, relationship~(\ref{eq:goal}) can also be expressed as
\begin{equation}\label{eq:goal2}
x_f^* = g\left(Fr,\Lambda_s\right).
\end{equation}
The determination of the functional dependencies (\ref{eq:goal}) and (\ref{eq:goal2}) is the primary objective of the present study.


Finally, note that in the configuration shown in figure \ref{fig:sketch}, the wall buoyancy flux, $\varphi$, is assumed to be uniform both beneath the gravity current and in the downstream free-convective region. In practice, this condition cannot be satisfied exactly because part of the electrical power supplied to the heaters is lost through the lower boundary and by radiation. As a result, the buoyancy flux beneath the current differs by about $15\%$ from that measured under free-convection conditions \citep{lanzini2026}. For simplicity, this difference is neglected, and the pre-release value of $\varphi$ is assumed to remain unchanged after the current is introduced.

\section{Experiments}\label{sec:exp}
\begin{figure}
\includegraphics[width=1.03\textwidth]{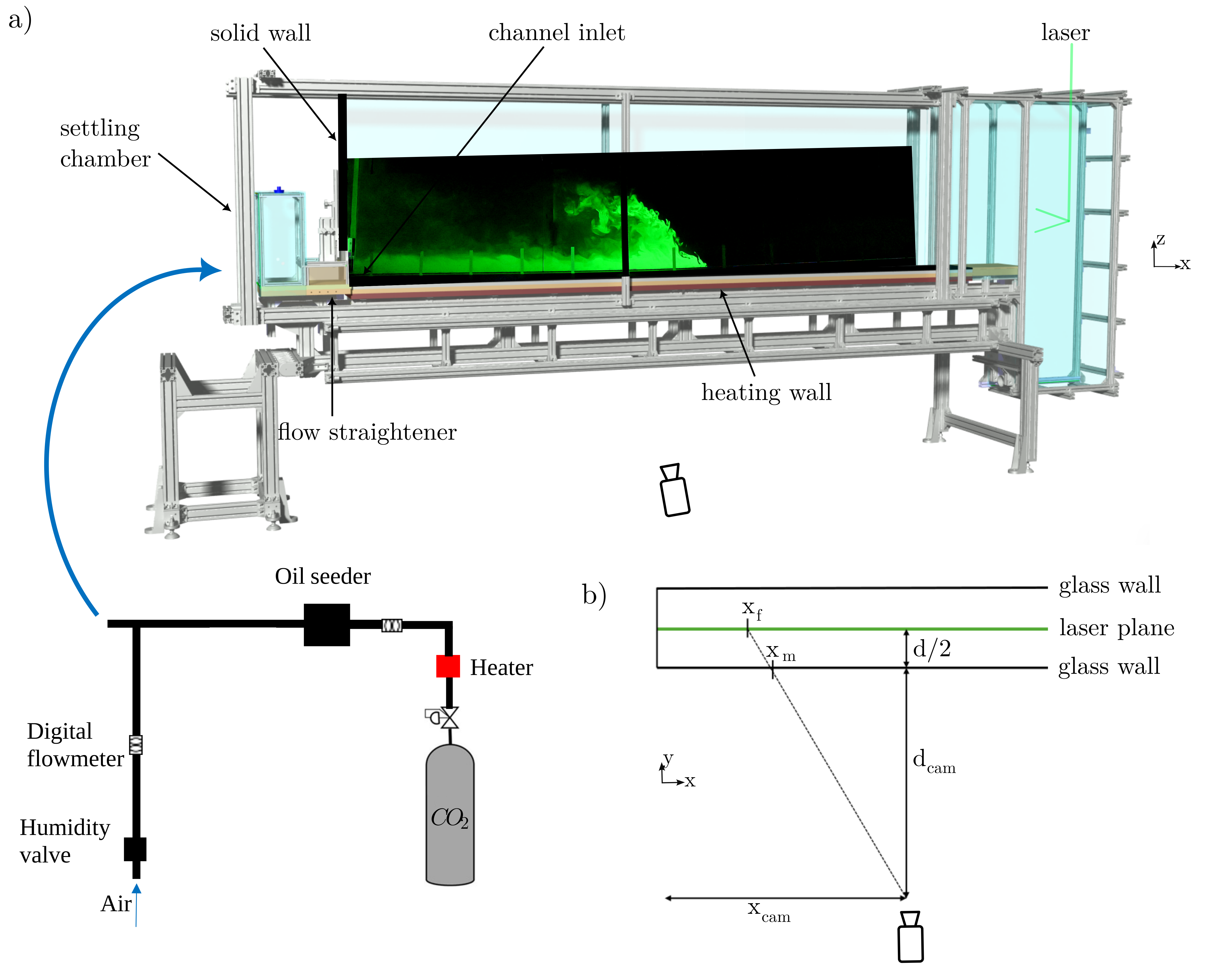}
    \caption{(a) Sketch of the experimental setup. (b) Perspective correction for front position measurements.}
    \label{fig:sketchexp}
\end{figure}
The experimental campaign is performed in a horizontal channel of length 
$l = 430\ \text{cm}$ and width $d = 30\ \text{cm}$, open at the top, whose bottom wall can be uniformly heated (figure \ref{fig:sketchexp}(a)). 
Details of the experimental facility, the heated wall, the measurement techniques used to determine the effective convective heat flux, $\varphi$, transferred to the overlying flow, and the assessment of its spatial homogeneity are given in \cite{lanzini2026}. Only the main features of the installation are therefore summarised here.
The first 400~cm of the bottom wall can be uniformly heated across the full channel width using eight heating resistive fabrics, each with a surface area of $500 \times 300$~mm$^2$, arranged adjacent to one another. All heating fabrics are powered in parallel at the same voltage. To homogenise the heat flux transferred to the overlying fluid, a 1~cm thick aluminium plate is placed above the heating elements. A series of insulating layers is installed beneath them to minimise downward heat losses. The total upward heat flux leaving the aluminium plate, $\varphi_{\text{meas}}$, is measured by means of heat-flux meters positioned above the plate at different streamwise distances $x$ from the source, together with the local wall temperature at the same locations. The convective heat flux is obtained by subtracting from the measured heat flux, $\varphi_{\text{meas}}$, the radiative contribution, $\varphi_r$, associated with thermal radiation exchanged between the heated floor and the surrounding walls of the room.
An uncertainty of $\pm 15\%$ is estimated for $\varphi$ \citep{lanzini2026}. 
Prior to the current injection, the bottom wall is heated uniformly for several hours until a steady thermal state of the heated wall is achieved.
The latter is assumed to be achieved when consecutive measurements of the wall temperature, taken at 10-minute intervals at a fixed point, show differences smaller than 2$\%$ (considering values in Celsius). Once this condition is achieved, the dense fluid is released into the channel.

The gravity currents are generated by a mixture of air and carbon dioxide injected at the channel inlet with a constant mass flux. Before entering the channel, the mixture is supplied in a $0.4$ m$^3$  settling chamber equipped with flow straighteners, consisting of a 10 cm long honeycomb with a circular cross-section of 0.6 cm diameter (figure \ref{fig:sketchexp}(a)). Once the settling chamber is filled, the inlet gate is opened, allowing the gravity current to propagate along the channel bed. The height of the inlet opening can be regulated between 1 and 10 cm. 
A relevant setup modification compared to the configuration described in \cite{lanzini2026} consists of adding a vertical solid wall above the channel inlet (figure \ref{fig:sketchexp}(a)). This additional barrier has a significant impact on the flow dynamics, as it suppresses the formation of a `sea-breeze' of external air that would otherwise arise due to the pure floor heating \citep{Townsend_1972} and enter the channel. In the absence of the barrier, the sea breeze would develop and superpose onto the injected gravity current front, thereby substantially altering the experimental results.
The dense current is seeded with micron-sized olive oil droplets and then visualized through laser tomography on a green light sheet coincident with the channel symmetry plane. 
The experimental run is recorded using a video camera. All videos are processed by tracking the front position as a function of time, by means of a reference grid placed on the front glass wall. The positions indicated by the grid are then corrected to account for perspective error (figure \ref{fig:sketchexp}(b)). The camera is placed at a distance $d_{cam}$ from the glass wall and at a horizontal distance $x_{cam}$ from the channel inlet. For a given position indicated by the grid, $x_{g}$, the corresponding front position is $x_{f}=x_g -(d/2)(x_{cam}-x_{g})/d_{cam}$ (figure \ref{fig:sketchexp}(b)). 
Flow visualizations do not provide quantitative CO$_2$ concentration measurements, which would require a pixel-wise calibration of image intensity \citep{Nogueira_2013,SherWoods2015}. They nevertheless provide valuable insight into the flow dynamics and enable robust identification of the front position, as the sharp concentration gradient at the current nose appears as a distinct brightness gradient in the images \citep{Nogueira_2013,SherWoods2017} (\S \ref{subsec:flowvisual}).

\begin{table} \centering \begin{tabular*}{0.8\textwidth}{@{\extracolsep{\fill}}ccccc} \hline $Re_s$ & $Fr_s$ & $\rho_s/\rho_0$ & $h_s$ [cm] & $u_s$ [m/s] \\ \hline 1640 & 1.5 & 1.052 & 8.0 & 0.29 \\ 1170 & 1.5 & 1.052 & 6.5 & 0.26 \\ 570 & 1.5 & 1.052 & 4.0 & 0.20 \\ 200 & 1.5 & 1.052 & 2.0 & 0.15 \\ 50 & 1.5 & 1.052 & 1.0 & 0.10 \\ \hline \end{tabular*} \caption{Preliminary adiabatic experiments ($\Lambda_s=0$) to investigate the influence of $Re_s$ on the front velocity.
} \label{tab:Re} 
\end{table}


Table \ref{tab:Re} summarises the experimental conditions for the preliminary adiabatic tests conducted to identify the threshold Reynolds number beyond which the front propagation becomes independent of $Re_s$. The results of these preliminary experiments, performed at a fixed source Froude number of $Fr_s = 1.5$, indicate that this independence is achieved for $Re_s \gtrsim 1000$ (see \S \ref{subsec:frontposition}). Accordingly, all subsequent non-adiabatic experiments are conducted at $Re_s \approx 1600$ (these are listed in Table \ref{tab:expcondition}). Four Froude numbers at the source, $Fr_s$, are tested. The same $Fr_s$ is achieved for different $f_s$ values, varying simultaneously $u_s$, $h_s$, and $\rho_s/\rho_0$ (the density at the source is changed by varying the CO$_2$ volume fraction in the mixture). Different values of $\Lambda_s$ are tested for each $Fr_s$ by varying both the heat flux (i.e., $f_\varphi$) and the inlet buoyancy flux, $f_s$.  The inlet conditions are labeled as `EXPIJ', where `I' is a number that goes from 1 to 4 corresponding to the different $Fr_s$ values, and `J' can be `a', `b', or `c' depending on the density ratio at the source. To allow a wider range of $\Lambda_s$ to be investigated experimentally, three density ratios are tested, i.e., $\rho_s/\rho_0=1.052$, $1.026$, and $1.015$ (Table \ref{tab:expcondition}), obtained with a CO$_2$ volume fraction at the source equal to $c_s=10 \%$, $5 \%$, and $3 \%$, respectively. Note that, in all cases, the relative density variations remain sufficiently small for the Boussinesq approximation to hold, thereby ruling out any direct influence of the density ratio on the flow dynamics  (see \S \ref{subsec:frontposition}).

To evaluate the experimental repeatability, at least six runs are repeated for each experimental configuration with $\Lambda_s<0.04$ and at least three for $\Lambda_s>0.04$. To recover the same ambient air statistical condition in the channel at the moment of the injection, 30 minutes are waited between two experimental repetitions. A discussion on the results of different repetitions of the same experiment is provided in \S \ref{subsec:frontposition}.
\begin{landscape}
\begin{table}
\centering
\resizebox{1.3\textwidth}{!}{
\begin{tabular}{cccccccc|ccccccccc}
\hline
name & $Fr_s$ & $Re_s$ & $\rho_s/\rho_0$ & $h_s$ [cm] & $u_s$ [m/s] & $\varphi$ [W/m$^2$] & $\Lambda_s$ 
& & name & $Fr_s$ & $Re_s$ & $\rho_s/\rho_0$ & $h_s$ [cm] & $u_s$ [m/s] & $\varphi$ [W/m$^2$] & $\Lambda_s$ \\ 
\hline
EXP1a & 3.0 & 1670 & 1.052 & 5.0  & 0.48 & 
\begin{tabular}[c]{@{}l@{}}0 \\ 30 \\ 100 \\ 180 \\ 240 \\ 300 \\ 360 \end{tabular} & 
\begin{tabular}[c]{@{}l@{}}0 \\ 0.003 \\ 0.010 \\ 0.020 \\ 0.025 \\ 0.030 \\ 0.040 \end{tabular}
& & 
EXP3a & 1.5 & 1640 & 1.052 & 8.0 & 0.29 & 
\begin{tabular}[c]{@{}l@{}}0 \\ 15 \\ 55 \\ 100 \\ 160 \\ 240 \\ 300 \\ 360 \end{tabular} &
\begin{tabular}[c]{@{}l@{}}0 \\ 0.003 \\ 0.010 \\ 0.020 \\ 0.030 \\ 0.045 \\ 0.060 \\ 0.070 \end{tabular} \\
EXP1b & 3.0 & 1640 & 1.026 & 6.5 & 0.38 & 
\begin{tabular}[c]{@{}l@{}}60 \\ 100 \\ 170 \\ 230\end{tabular} & 
\begin{tabular}[c]{@{}l@{}}0.017 \\ 0.025 \\ 0.050 \\ 0.070\end{tabular} 
& & 
EXP3b & 1.5 & 1600 & 1.026 & 10 & 0.24 &
\begin{tabular}[c]{@{}l@{}}\\130 \\ 160 \\ 220 \\ 300\end{tabular} & 
\begin{tabular}[c]{@{}l@{}}\\0.060 \\ 0.075 \\ 0.105 \\ 0.140\end{tabular} \\

EXP1c & 3.0 & 1600 & 1.015 & 7.5 & 0.315 & 
\begin{tabular}[c]{@{}l@{}}\\26 \\ 43 \\ 70 \\ 120 \\ 170 \\ 220\end{tabular} &
\begin{tabular}[c]{@{}l@{}}\\0.015 \\ 0.025 \\ 0.040 \\ 0.070 \\ 0.100\\ 0.130\end{tabular} 
& & & & & & & & & \\
\hline
EXP2a & 2.0 & 1570 & 1.052 & 6.3 & 0.36 & 
\begin{tabular}[c]{@{}l@{}}0 \\ 20 \\ 65 \\ 125 \\ 210 \\ 240 \\ 300\end{tabular} &
\begin{tabular}[c]{@{}l@{}}0 \\ 0.003 \\ 0.010 \\ 0.020 \\ 0.030 \\ 0.040 \\ 0.045\end{tabular} 
& & 
EXP4a & 0.9 & 1480 & 1.052 & 10 & 0.21 & 
\begin{tabular}[c]{@{}l@{}}0 \\ 12 \\ 34 \\ 65 \\ 110 \\ 240 \\ 300 \\ 360 \end{tabular} &
\begin{tabular}[c]{@{}l@{}}0 \\ 0.003 \\ 0.010 \\ 0.020 \\ 0.030 \\ 0.065 \\ 0.080 \\ 0.095 \end{tabular} \\
EXP2b & 2.0 & 1740 & 1.026 & 8.5 & 0.30 & 
\begin{tabular}[c]{@{}l@{}}170 \\ 230\end{tabular} &
\begin{tabular}[c]{@{}l@{}}0.060 \\ 0.085\end{tabular}
& & & & & & & & & \\
EXP2c & 2.0 & 1580 & 1.015 & 10 & 0.25 & 160 & 0.120 & & & & & & & & & \\

\end{tabular}}
\caption{Tested turbulent experimental conditions.}
\label{tab:expcondition}
\end{table}
\end{landscape}
\section{Results}
\label{sec:results}

\subsection{Flow visualizations}\label{subsec:flowvisual}
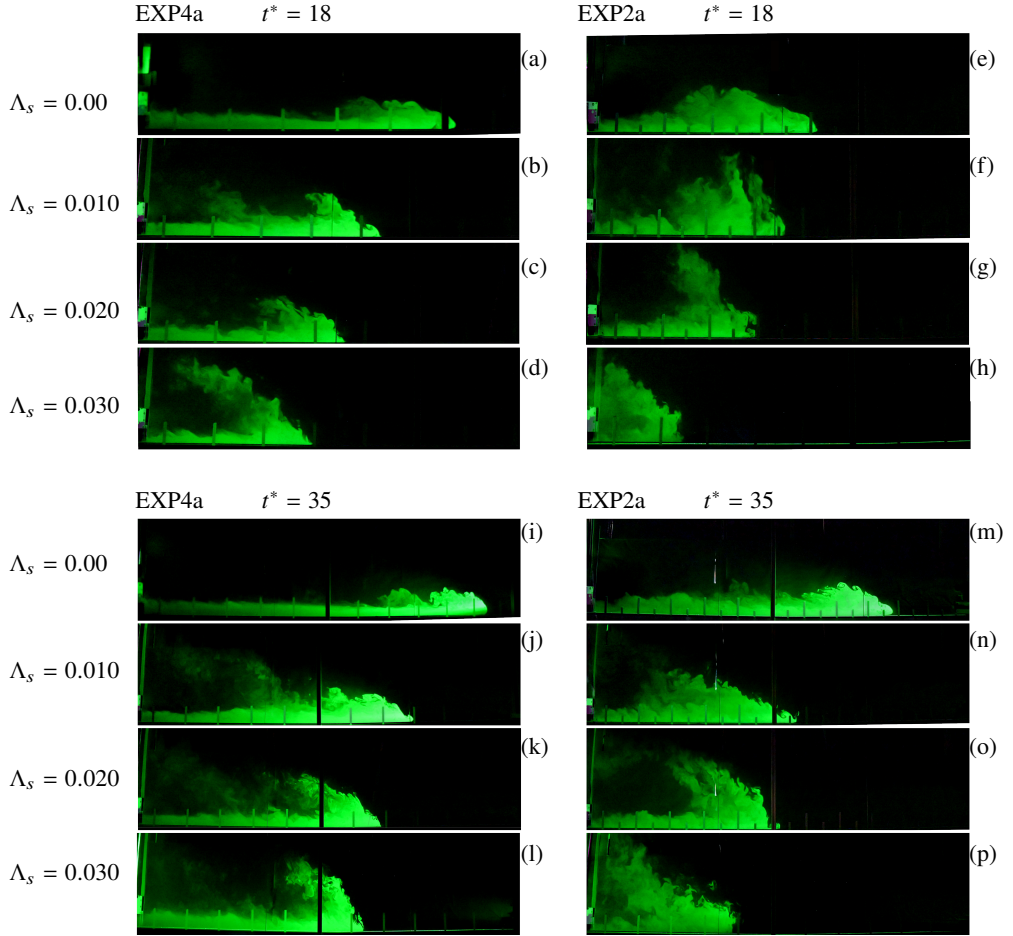
\begin{figure}
\centering
  \unitlength = 0.835cm
\begin{picture}(15,15.5)
\put(-0.5,0){\includegraphics[width=1.05\textwidth]{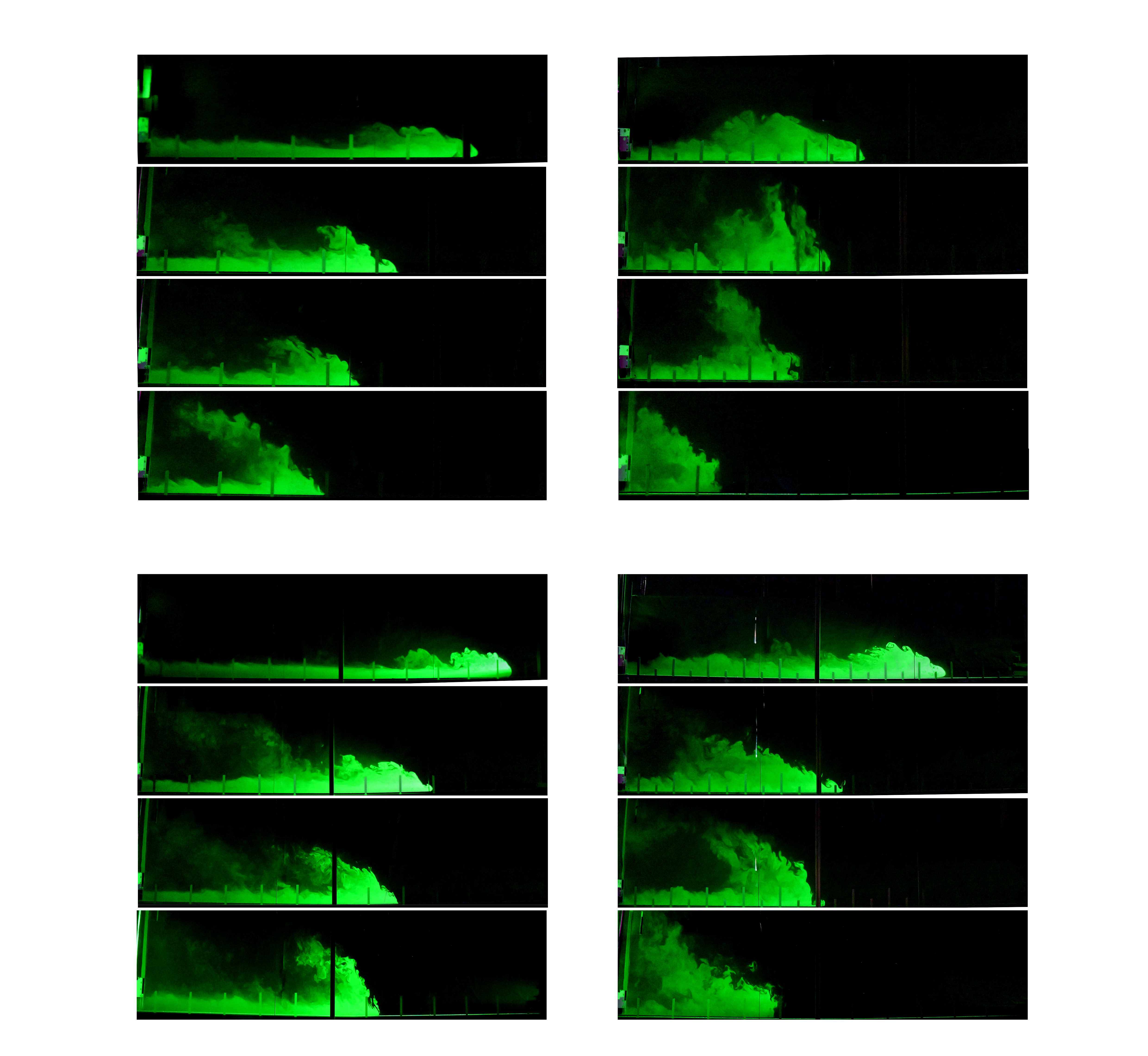}}
\put(-0.5,1.4){$\Lambda_s=0.030$}
\put(-0.5,2.9){$\Lambda_s=0.020$}
\put(-0.5,4.6){$\Lambda_s=0.010$}
\put(-0.5,6.3){$\Lambda_s=0.00$}
\put(-0.5,8.8){$\Lambda_s=0.030$}
\put(-0.5,10.3){$\Lambda_s=0.020$}
\put(-0.5,12){$\Lambda_s=0.010$}
\put(-0.5,13.6){$\Lambda_s=0.00$}
\put(1.5,7.3){EXP4a}
\put(8.5,7.3){EXP2a}
\put(3.5,7.3){$t^*=35$}
\put(10.5,7.3){$t^*=35$}
\put(1.5,15){EXP4a}
\put(8.5,15){EXP2a}
\put(3.5,15){$t^*=18$}
\put(10.5,15){$t^*=18$}
\put(7.6,14.3){(a)}
\put(7.6,12.6){(b)}
\put(7.6,11){(c)}
\put(7.6,9.4){(d)}
\put(14.7,14.3){(e)}
\put(14.7,12.6){(f)}
\put(14.7,11){(g)}
\put(14.7,9.4){(h)}

\put(7.6,6.8){(i)}
\put(7.6,5.1){(j)}
\put(7.6,3.4){(k)}
\put(7.6,1.7){(l)}
\put(14.7,6.8){(m)}
\put(14.7,5.1){(n)}
\put(14.7,3.4){(o)}
\put(14.7,1.7){(p)}
\end{picture}
    \caption{Flow visualization of EXP4a ($Fr_s=0.9$) ((a)-(d), (i)-(l)) and EXP2a ($Fr_s=2.0$) ((e)-(h),(m)-(p)) currents for $\Lambda_s=0$, $\Lambda_s=0.010$, $\Lambda_s=0.020$, and $\Lambda_s=0.030$. The snapshots are taken at two different non-dimensional times: $t^*=18$ ((a)-(h)) and $t^*=35$ ((i)-(p)). }
    \label{fig:visual}
\end{figure}
Our analysis begins with a phenomenological description of the current based on flow visualizations. Figure \ref{fig:visual} compares the EXP4a and EXP2a (with $Fr_s=0.9$ and $Fr_s=2.0$, respectively (Table\ref{tab:expcondition})) currents at two different non-dimensional times, $t^*=18$ (Fig. \ref{fig:visual}(a)-(h)) and $t^*=35$ (Fig. \ref{fig:visual}(i)-(p)), obtained for $\Lambda_s=0$, $0.010$, $0.020$, and $0.030$. For both EXP4a and EXP2a, the heating induces a deceleration in the front velocity and modifies the current structure. The slowdown is stronger as $\Lambda_s$ increases.
Despite this, similar to the non-heated case, at a sufficient distance from the source, it is always possible to identify, at the leading edge, a thicker head characterised by intense mixing followed by the shallower current body through which the mixing is weaker.

The reduction of the front speed can be attributed to two distinct physical mechanisms.
The first is related to the direct heating of the gravity-current fluid. As the current propagates along the wall, heating progressively reduces the head integral buoyancy, $b_hh_h$, relative to the adiabatic case (see \S\ref{sec:ch3model}). Under the assumption that heating does not affect the value of head Froude number $Fr$, the resulting reduction in $b_hh_h$ directly translates into a decrease in the front velocity, $\mathrm{d}x_f/\mathrm{d}t$ (Eq. \ref{eq:Fr}), with respect to the non-heated case.
The second physical mechanism that causes the slowdown of the front involves the interaction between the horizontal flow and the vertical thermal plumes induced by free convection downstream of the current front. 
Although the intensity and position of the thermals are intermittent, they exert a clear influence on the advancing current (figure \ref{fig:visual}). 
When the front encounters a thermal plume, a highly unstable flow condition is created as the warm ambient air mixes with the denser fluid at the head, resulting in a significant increase in turbulent intensity and mixing, together with a rapid rise in the front height.
This interaction increases the resistance to the front horizontal propagation. Enhanced drag has also recently been measured in the body of heated gravity currents (i.e., in the region upstream of the head) relative to the adiabatic condition \citep{lanzini2026}. The increased drag acting on the head is associated with a reduction in the head Froude number, $Fr$. Such a reduction is systematically observed, for example, in adiabatic gravity currents propagating over surfaces with increasing roughness, i.e., with enhanced resistance to the horizontal motion with respect to the smooth case \citep{Nogueira2013,Maggi2022, maggi2025}.
An estimate of the respective contributions of the two physical mechanisms responsible for the front deceleration, namely the progressive heating of the gravity-current fluid and the enhanced drag, is obtained from the model results presented in \S \ref{sec:ch3model}.

\begin{figure}
\centering
\unitlength = 1.0cm
\begin{picture}(15,4)
\put(0.2,0){\includegraphics[height=4cm]{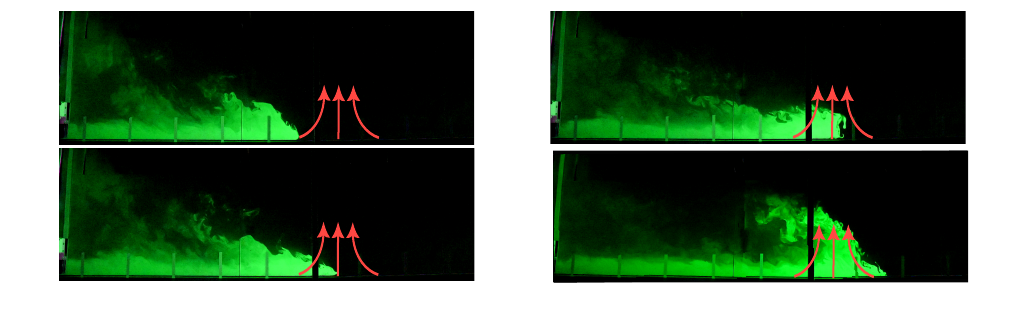}}
\put(0.45,3.5){(a)}
\put(0.45,1.7){(b)}
\put(6.8,3.5){(c)}
\put(6.8,1.7){(d)}
\end{picture}
  \caption{Sequence of snapshots showing the interaction between the current front and a thermal plume (shown by the red arrows) in an EXP4a experiment with $\Lambda_s=0.030$. Snapshots (a)-(d) are taken 12.9, 13.5, 14.1, and 15.7 seconds after the current injection, respectively.}
  \label{fig:visual2}
\end{figure}

A detailed sequence of the interaction phases between the current and a thermal plume for an EXP4a current with $\Lambda_s=0.030$ is shown in figure \ref{fig:visual2}. The snapshot in figure \ref{fig:visual2}(a) is taken 12.9 seconds after the gate opening. At 13.5 seconds after the gate opening (figure \ref{fig:visual2}(b)), the front thins and accelerates. This acceleration can be attributed to the presence of the hot air column immediately downstream of the front, which induces a temporary increase in the static pressure difference between the front and the surrounding environment, resulting in an acceleration. The angle identified by the head edge is reduced compared to that shown in figure \ref{fig:visual2}(a). Subsequently, 14.1 seconds after the injection (figure \ref{fig:visual2}(c)), the front encounters the column of hot air moving upward, causing a deceleration in the front speed and an increase in the head height. In this snapshot, the angle formed by the head edge with the bottom wall is approximately 90 degrees.   The head height continues to increase until the front overcomes the plume (figure \ref{fig:visual2}(d)). In this case, a small structure forms near the wall, which begins to move forward, leaving the plume behind.  The hot plume entrained into the head causes a reduction of the fluid density within the head. Differently from the non-heated case, the head fluid does not settle back directly into the current body \citep{Simpson_Britter_1979, SherWoods2017}, but rather remains suspended and produces a wake behind the head that is considerably longer with respect to the adiabatic case (figure \ref{fig:visual}).

\subsection{Front position}\label{subsec:frontposition}
\begin{figure}
\centering
 \unitlength = 1.0cm
\begin{picture}(10,5.1)(2,0)
\includegraphics[height=5.1cm]{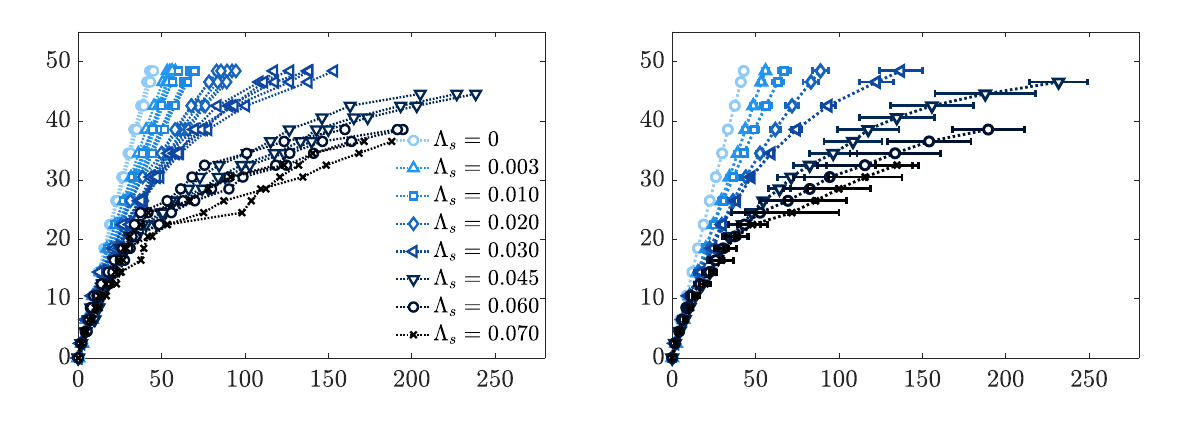}
\put(-13.1,4.4){(a)}
\put(-6.1,4.4){(b)}
\put(-14,2.55){$x_f^*$}
\put(-7,2.55){$x_f^*$}
\put(-10.7,0.2){$t^*$}
\put(-3.7,0.2){$t^*$}
\end{picture}
\caption{Evaluation of the experimental repeatability for the EXP3a source condition. Panel (a) shows the non-dimensional front position for all the experimental repetitions for EXP3a, while panel (b) shows their ensamble average and the dispersion bars equal to $\sigma/\sqrt{N}$.}
    \label{fig:repetition}
\end{figure}
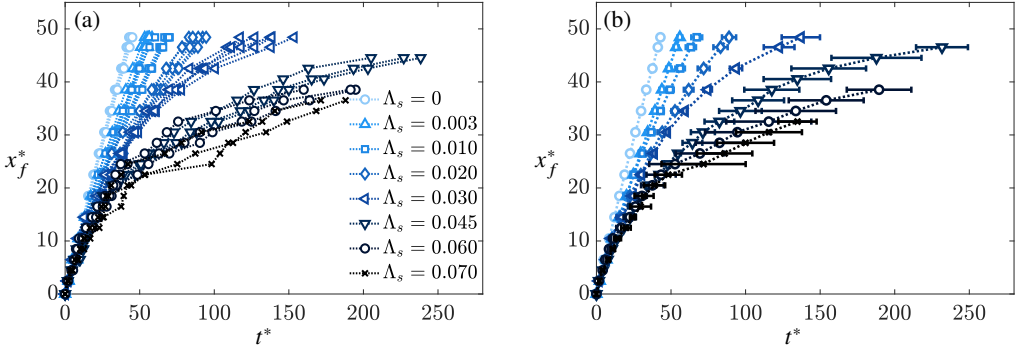

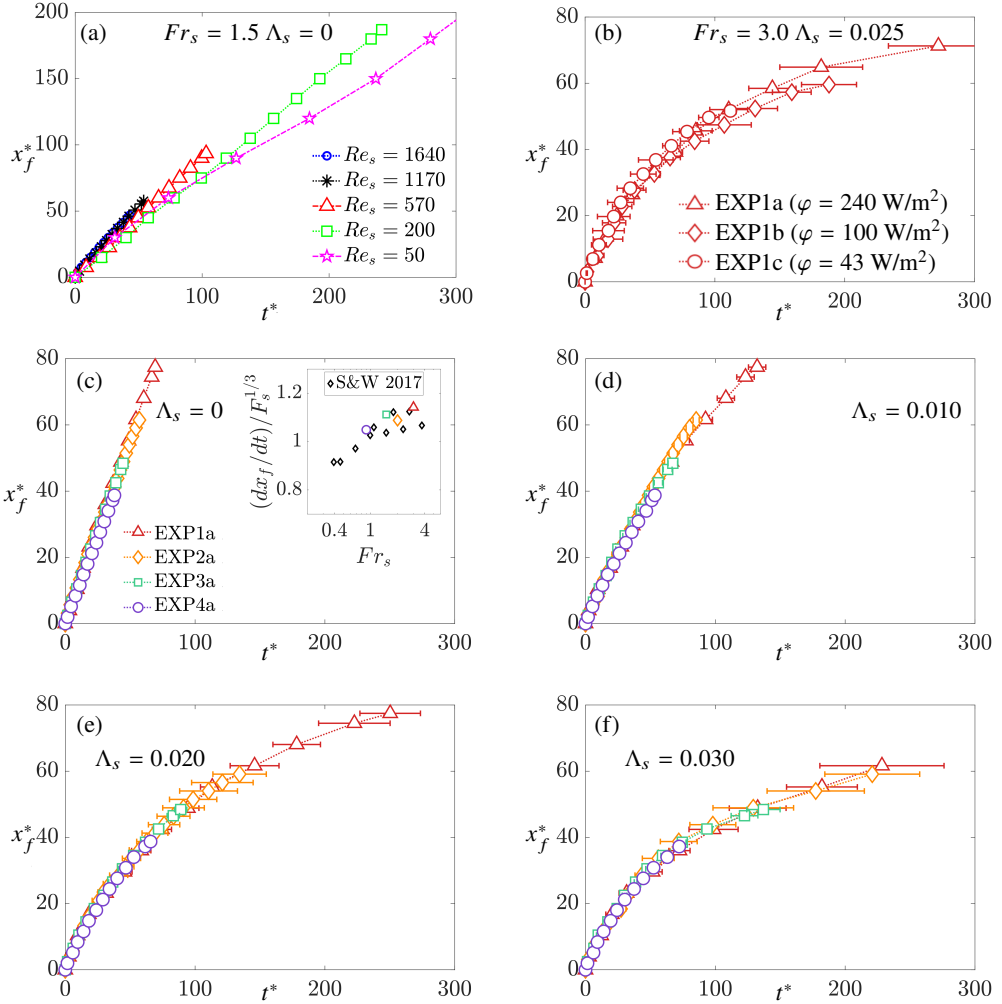
\begin{figure}
\centering
 \unitlength = 1.0cm
\begin{picture}(10,14)(1.7,0)
\includegraphics[height=14cm]{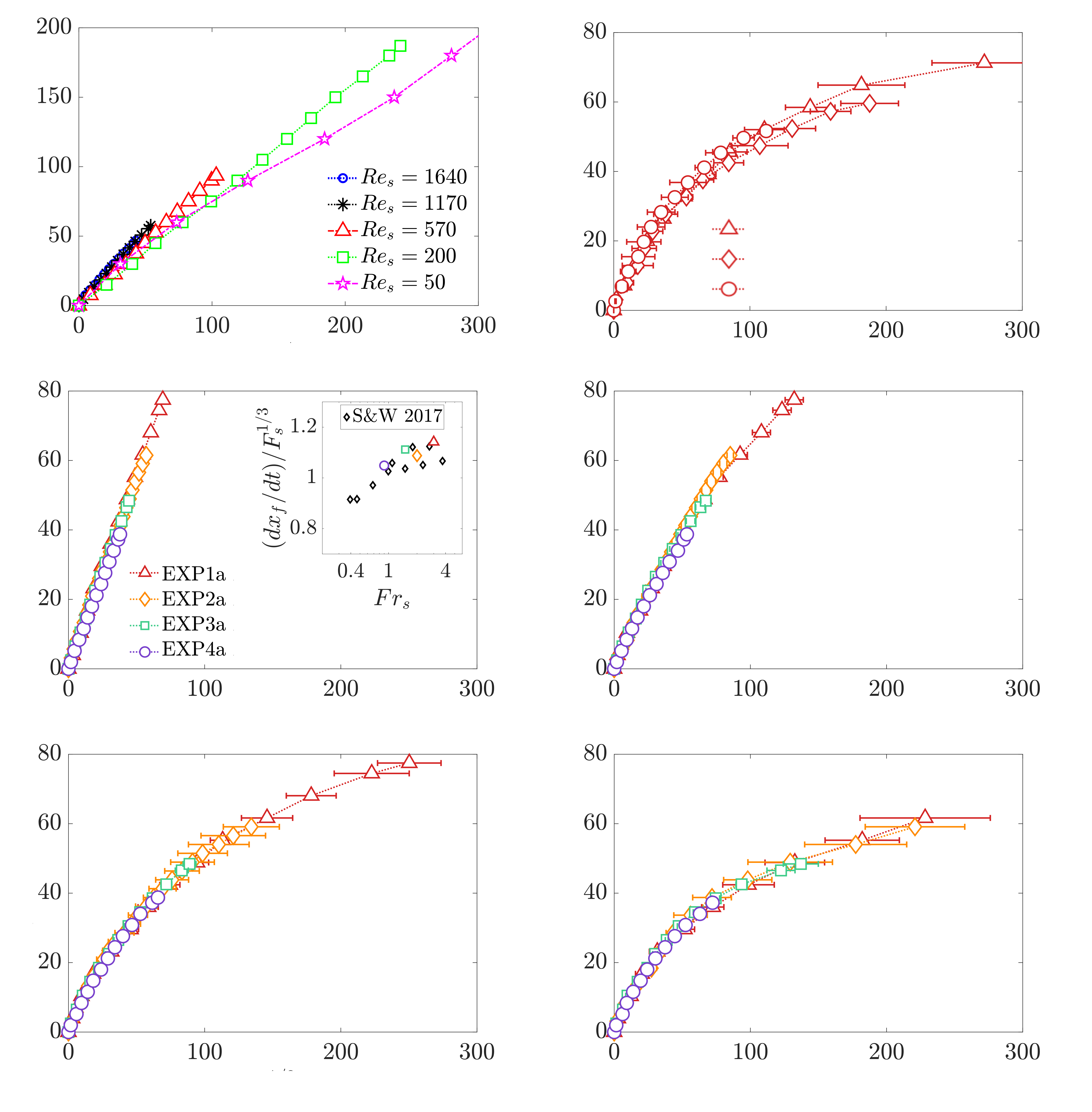}
\put(-4.3,11.05){EXP1a ($\varphi=240$ W/m$^2$)}
\put(-4.3,10.65){EXP1b ($\varphi=100$ W/m$^2$)}
\put(-4.3,10.25){EXP1c ($\varphi=43$ W/m$^2$)}
\put(-11.7,8.3){$\Lambda_s=0$}
\put(-11.6,13.3){$Fr_s=1.5$ $\Lambda_s=0$}
\put(-4.6,13.3){$Fr_s=3.0$ $\Lambda_s=0.025$}
\put(-2.5,8.3){$\Lambda_s=0.010$}
\put(-12.5,3.7){$\Lambda_s=0.020$}
\put(-5.5,3.7){$\Lambda_s=0.030$}
\put(-12.7,13.3){(a)}
\put(-5.9,13.3){(b)}
\put(-12.75,8.7){(c)}
\put(-5.9,8.7){(d)}
\put(-12.75,4.15){(e)}
\put(-5.9,4.15){(f)}
\put(-6.8,11.7){$x^*_f$}
\put(-13.6,11.7){$x^*_f$}
\put(-6.8,7.2){$x^*_f$}
\put(-13.7,7.2){$x^*_f$}
\put(-6.8,2.7){$x^*_f$}
\put(-13.6,2.7){$x^*_f$}
\put(-10.3,9.6){$t^*$}
\put(-3.5,9.6){$t^*$}
\put(-10.3,5.1){$t^*$}
\put(-3.5,5.1){$t^*$}
\put(-10.3,0.6){$t^*$}
\put(-3.5,0.6){$t^*$}
\end{picture}
\caption{Panel (a) shows the influence of $Re_s$ on $x^*_f$ for the adiabatic case, and panel (b) shows a comparison between the turbulent experimental results obtained for fixed $\Lambda_s$ and $Fr_s$. Panels (c)-(f) show the results with $\Lambda_s=0$, $\Lambda_s=0.010$, $\Lambda_s=0.020$, and $\Lambda_s=0.030$, respectively, for different $Fr_s$. The subplot of panel (b) shows a comparison between the non-dimensional constant front velocity measured in this work for $\Lambda_s=0$ and the experimental results of \cite{SherWoods2017}.}
    \label{fig:similarity}
\end{figure}
\begin{figure}
\centering
 \unitlength = .835cm
\begin{picture}(10,18.)(3,0)
\includegraphics[width=1\textwidth]{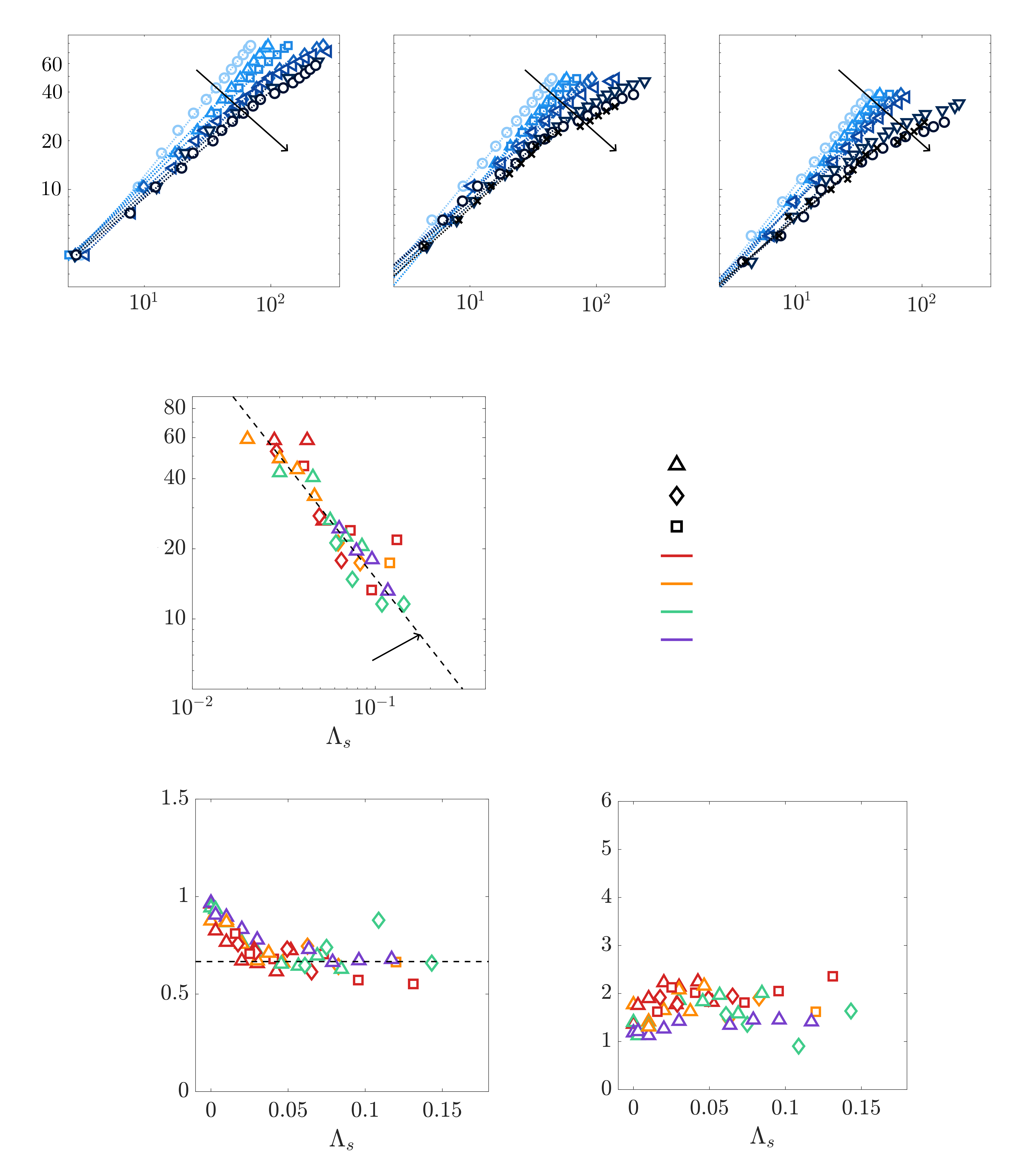}
\put(-5.15,11.0){${\rho_s}/{\rho_0}=1.052$}
\put(-5.15,10.5){${\rho_s}/{\rho_0}=1.026$}
\put(-5.15,10.){${\rho_s}/{\rho_0}=1.015$}
\put(-5.15,9.5){$Fr_s=3.0$}
\put(-5.15,9.05){$Fr_s=2.0$}
\put(-5.15,8.6){$Fr_s=1.5$}
\put(-5.15,8.15){$Fr_s=0.9$}
\put(-15,17.4){(a)}
\put(-9.9,17.4){(b)}
\put(-4.7,17.4){(c)}
\put(-13,11.7){(d)}
\put(-13,5.4){(e)}
\put(-6.3,5.4){(f)}
\put(-14,3.4){$A$}
\put(-7.3,3.4){$B$}
\put(-14.8,16.6){EXP1a}
\put(-9.6,16.6){EXP3a}
\put(-4.5,16.6){EXP4a}
\put(-16.2,15.7){$x^*_f$}
\put(-13,13.2){$t^*$}
\put(-8,13.2){$t^*$}
\put(-2.7,13.2){$t^*$}
\put(-12,15.6){$\Lambda_s$}
\put(-6.8,15.6){$\Lambda_s$}
\put(-2,15.6){$\Lambda_s$}
\put(-14.2,9.7){$x_c^*$}
\put(-10.9,7.9){$\frac{C_1}{\Lambda_s}$}
\end{picture}
\caption{Panels (a),(b),(c) show the non-dimensional front position as a function of the non-dimensional time for all the tested $\Lambda_s$ with EXP1a (a), EXP3a (b), and, EXP4a (c). Panel (d) shows the distance from the source up to which the power-law fitting of the results has an $R^2>0.975$ ($C_1=1.5$). Panel (e) and (f) show the coefficients of the power-law fitting (\ref{eq:powerlaw}), for $x^*_f<x_c^*$.}
    \label{fig:loglog}
\end{figure}
We now turn to a quantitative analysis of the front propagation, with the objective of investigating the relation (\ref{eq:goal}).
Figures \ref{fig:visual} and \ref{fig:visual2} show that the instantaneous velocity of the front depends on its interaction with vertical convective structures induced by natural convection, which do not have a fixed position and intensity, given the turbulent nature of the phenomenon. 
As a result, the experimental outcomes of a single run are influenced by the instantaneous flow field in the channel when the current is generated. To characterise the run-to-run variability, each test was repeated six times for $\Lambda_s \leq 0.040$ and at least three times for $\Lambda_s > 0.040$. Figure~\ref{fig:repetition}(a) shows all the repetitions with the inlet condition EXP3a for different heating levels. The dispersion between the repetitions increases with $\Lambda_s$, reflecting the enhanced intensity of the thermal plumes and their stronger influence on the front propagation.
In figure \ref{fig:repetition}(b) and in the remainder of the manuscript, only the results obtained from the ensemble average of the different repetitions are presented, together with horizontal dispersion bars equal to $\sigma/\sqrt{N}$, where $\sigma$ is the standard deviation of the times $t^*$ recorded at the passage of the front at a given position $x^*_f$, and $N$ is the number of experimental repetitions of the same test \citep{MOFFAT1988}.

The results of the preliminary adiabatic experiments (Table \ref{tab:Re}), performed to assess the influence of the Reynolds number on front propagation, are shown in figure \ref{fig:similarity}(a). For $Re_s=50$ and $Re_s=200$, the front decelerates as it moves downstream from the source. A similar behaviour was reported by \cite{Britter_Linden_1980} in experiments on low-Reynolds-number sustained gravity currents. At higher $Re_s$, the front advances at an approximately constant speed, which becomes effectively independent of $Re_s$ for $Re_s \gtrsim 1000$. This justifies the choice of $Re_s \approx 1600$ for all subsequent turbulent experiments.
A verification of the correct similarity parameter choice in the turbulent case -- that is, a demonstration that, for fixed governing parameters in equation (\ref{eq:goal}), the front position is uniquely determined -- is provided in figure~\ref{fig:similarity}(b), showing three experimental results obtained for the inlet conditions EXP1a, EXP1b, and EXP1c (i.e., at a fixed $Fr_s=3.0$) and at a fixed $\Lambda_s=0.025$. Both the source buoyancy flux $f_s$ and the heating $\varphi$ vary among the three experiments, with the heating in case EXP1a being almost six times larger than in EXP1c. Nevertheless, when appropriately rescaled, the experimental results collapse onto a single curve (within experimental dispersion bars), thereby confirming the appropriateness of the chosen similarity parameters.
Figure \ref{fig:similarity}(c)-(f) compares the front position as a function of time for currents with varying source Froude numbers (EXP1a, EXP2a, EXP3a, and EXP4a) at $\Lambda_s=0,0.010,0.020,0.030$, respectively. Figure \ref{fig:similarity}(c) presents the results for the adiabatic experiments. Consistently with literature results on non-heated turbulent sustained gravity currents advancing along a horizontal wall, the front velocity is constant and slightly increases with $Fr_s$ \citep{Hoggetal2005,Hoggetal2016,SherWoods2017}. In the subplot of figure \ref{fig:similarity}(c), good agreement is shown between the non-heated constant front velocity obtained in this work and in the experiments of \citet{SherWoods2017}. For the non-adiabatic cases (figure \ref{fig:similarity}(d)-(f)), the curves overlap for each of the $\Lambda_s$ values considered, indicating that, within the studied range of $Fr_s$,  the source Froude number has weak influence on the front speed in the heated cases.

Figure \ref{fig:loglog}(a)-(c) shows the front position results, obtained for all the tested $\Lambda_s$, for the EXP1a, EXP3a, and EXP4a cases (Table \ref{tab:expcondition}). 
Notably, the experiments with small $\Lambda_s$ follow a linear trend on a double-logarithmic scale along the entire channel length. For higher $\Lambda_s$ values, this linearity holds only up to a critical distance from the source, $x_{c}^*=x_{c}/h_s$, after which the deceleration of the front is stronger. Therefore, for $x^*_f<x_{c}^*$, it is possible to write the non-dimensional front position, $x^*_f=x_f(t)/h_s$,  as:
\begin{equation}\label{eq:powerlaw}
    x^*_f=B(Fr_s,\Lambda_s)t^{*A(Fr_s,\Lambda_s)},
\end{equation}
which provides a more explicit functional representation of the relationship given in Eq. (\ref{eq:goal}).
Figure \ref{fig:loglog}(d) shows the values of the critical position $x_{c}^*$ up to which the power-law behaviour (Eq. \ref{eq:powerlaw}) holds as a function of $\Lambda_s$. These are obtained by applying a least-squares linear regression of the log-log results and considering the distance from the inlet up to where the coefficient of determination $R^2$ of the fit is higher than 0.975. Different marker colors correspond to different $Fr_s$, and different shapes correspond to different density ratios at the source.
An analogous approach is employed to identify the transition from the slumping and self-similar phases in lock-exchange gravity currents in \cite{Nogueira2013}.
The values of $x_c^*$, irrespective of the source Froude number, are well represented by the relation $C_1 / \Lambda_s$. The numerical value of $C_1$ depends on the arbitrary choice of the $R^2$ threshold adopted to define the validity range of the power-law, but, for $0.9 <R^2 < 0.975$, it was verified that the value of $x_c^*$ is always reliably captured by a $C_1 / \Lambda_s$ law.


Figure~\ref{fig:loglog}(e,f) reports the coefficients $A$ and $B$ of the power-law fit~\eqref{eq:powerlaw}, for $x^*_f < x_c^*$ and minimum accepted $R^2=0.975$ ($C_1=1.5$). In the non-heated case, the power-law exponent $A$ is close to unity, in agreement with figure~\ref{fig:similarity}(a). In the range $0<\Lambda_s<0.04$, $A$ decreases approximately linearly with $\Lambda_s$, for all the tested $Fr_s$. For $\Lambda_s>0.04$, the values of $A$ scatter around a constant value of $2/3$: this exponent coincides with the minimum slope attained within the power-law region in figure \ref{fig:loglog}(a)-(c). The values of the $B$ coefficient are more dispersed and have values in the range from 1 to 2.

\subsection{Stop position}
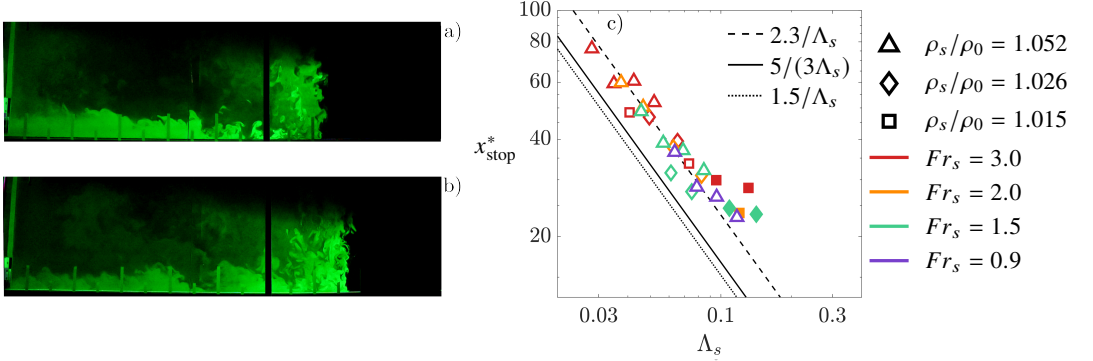
\begin{figure}
\centering
 \unitlength = 1.0cm
\begin{picture}(14,5.5)(0.4,0)
\includegraphics[width=1\textwidth]{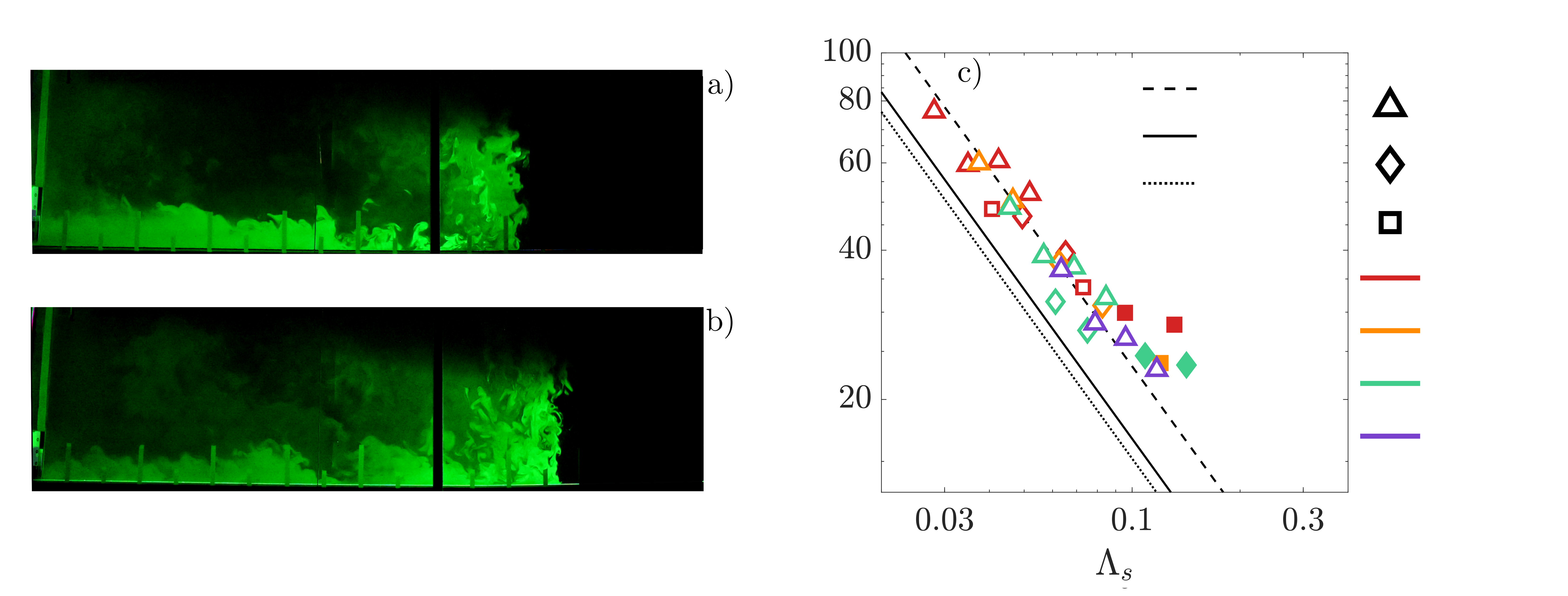}
\put(-1.05,4.15){${\rho_s}/{\rho_0}=1.052$}
\put(-1.05,3.65){${\rho_s}/{\rho_0}=1.026$}
\put(-1.05,3.15){${\rho_s}/{\rho_0}=1.015$}
\put(-1.05,2.63){$Fr_s=3.0$}
\put(-1.05,2.18){$Fr_s=2.0$}
\put(-1.05,1.73){$Fr_s=1.5$}
\put(-1.05,1.28){$Fr_s=0.9$}
\
\put(-3.1,4.25){$2.3/\Lambda_s$}
\put(-3.1,3.85){$5/(3\Lambda_s)$}
\put(-3.1,3.45){$1.5/\Lambda_s$}
\put(-7,2.8){$x_{\text{stop}}^{*}$}
\end{picture}
\caption{Panels (a) and (b) show snapshots of a strongly heated current with a null advancing front velocity: (a) is obtained for EXP4a with $\Lambda_s=0.095$ and (b) for EXP2a with $\Lambda_s=0.045$. (c) Experimental stop position $x_{\text{stop}}^{*}$ as a function of $\Lambda_s$. The filled markers show the experimental points that start to deviate from the $2.3/\Lambda_s$ law. The curves $5/(3\Lambda_s)$ and $1.5/\Lambda_s$ denote the stopping positions predicted by the model (\S \ref{sec:ch3model}). }
    \label{fig:stop}
\end{figure}

In experiments with sufficiently strong heating, the front eventually came to a halt. As an example, two snapshots of a stopped front, taken from the experiments with $Fr_s=0.9$ and $\Lambda_s=0.21$ (EXP4a), and $Fr_s=2.0$ and $\Lambda_s=0.11$ (EXP2a), are shown in figures \ref{fig:stop}(a) and \ref{fig:stop}(b), respectively.  In this situation, the heating of the current fluid is so intense that the density difference between the head and the ambient fluid immediately downstream is cancelled (see \S \ref{sec:ch3model}). When this occurs, a statistically steady flow is created: the front no longer advances horizontally, and the height of the head begins to increase indefinitely. Since the density in the head is lower than the density far from the wall, $\rho_0$, a positively buoyant plume is generated at the head. The current body continuously supplies the air-carbon dioxide mixture to the head to maintain the plume formation. For all the $Fr_s$ investigated, it was possible, with sufficiently intense heating, to halt the front propagation before the end of the channel. The obtained non-dimensional stop positions, $x_{\text{stop}}^{*}=x_{\text{stop}}/h_s$, are shown in figure \ref{fig:stop}(c). Due to turbulent motions, the experimental stop position exhibits oscillations. We estimate an uncertainty of $\pm 15\,$cm associated with the determination of $x_{\text{stop}}$.
The experimental data are accurately fitted by the expression $x_{\text{stop}}^{*}=2.3/\Lambda_s$, and do not show a clear dependence on $Fr_s$. 

Only a few experiments with the largest heating rates ($\Lambda_s>0.1$, filled markers in figure \ref{fig:stop}(c)) deviate from the $2.3/\Lambda_s$ scaling. In these cases, the stop position is sufficiently close to the source that it may still lie within the hydraulic adjustment region -- whose extent increases as $Fr_s$ increases \citep{SherWoods2017} -- where the flow is still adjusting from the source conditions toward the head-controlled Froude number \citep{SherWoods2017,Horsley2018,Ungarish2023,Harrouk_Mehaddi_Arcen_Dossmann_2025}.
In the hydraulic adjustment region, the current properties are still strongly affected by the source Froude number.
Indeed, for the experiments identified by filled markers in figure \ref{fig:stop}(c), the arrest distance increases when $Fr_s$ is reduced, i.e., when the current is more forced at the source.

\section{Analytical model for the front position}\label{sec:ch3model}
\begin{figure}
\centering
  \unitlength = 1.cm
\begin{picture}(12,13.5)
\put(-1.0,0){\includegraphics[height=13.5cm]{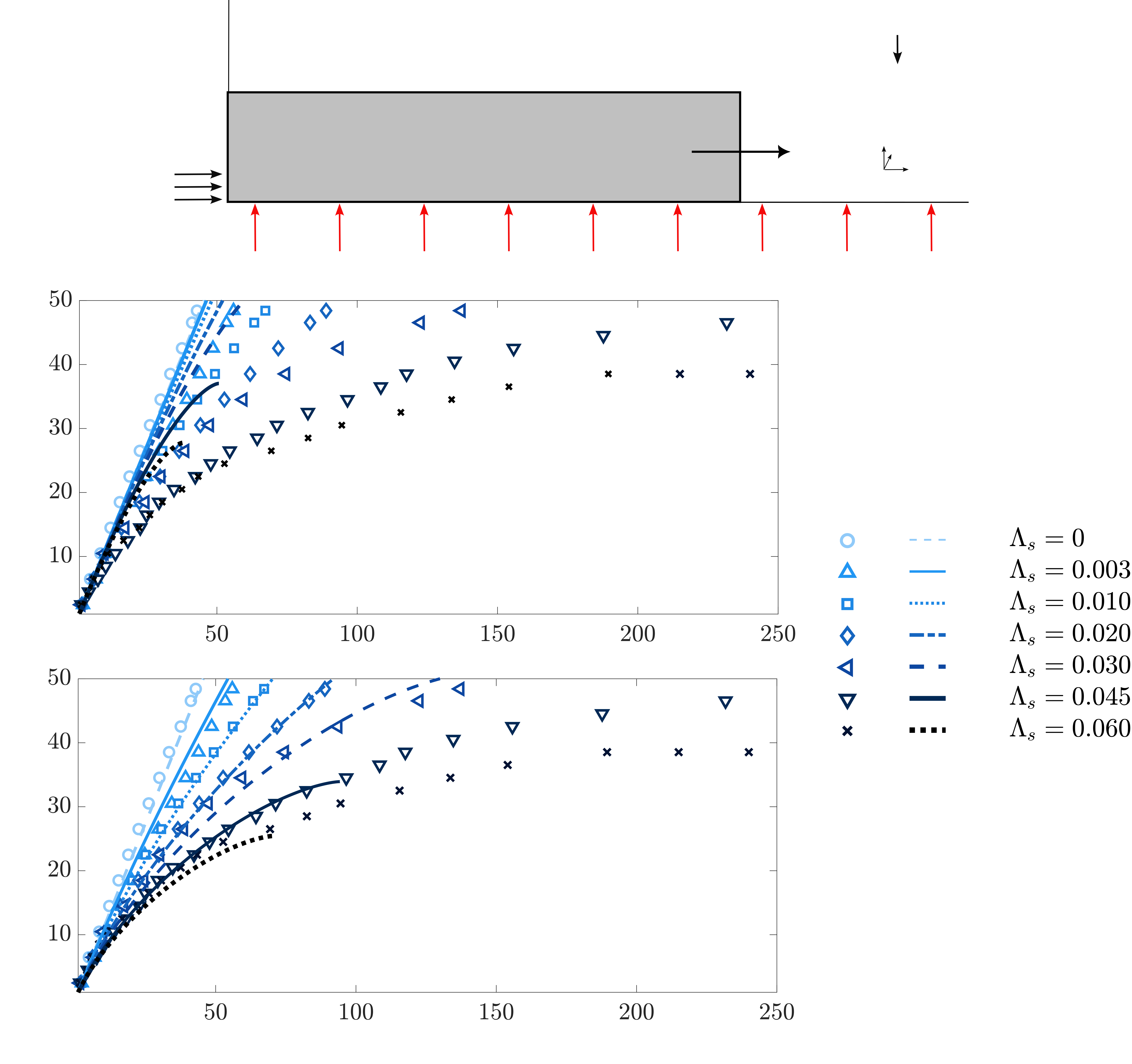}}
\put(1.4,13.2){a)}
\put(0.05,9.4){b)}
\put(0.0,4.6){c)}
\put(-0.8,7.7){$x_f^*$}
\put(-0.8,2.9){$x_f^*$}
\put(4.5,0.3){$t^*$}
\put(0.6,12.25){$T_s=T_0$}
\put(1.3,12.0){$c_s$}
\put(1.3,11.7){$u_s$}
\put(1.3,11.4){$h_s$}
\put(6.9,12.8){$T_0$,  $\rho_0$}
\put(8.5,11.2){$x_f(t)$}
\put(10,12.8){$g$}
\put(10,11.7){$z$}
\put(10.5,11.2){$x$}
\put(10.3,11.6){$y$}
\put(6.1,10.4){\textcolor{red}{$\varphi$}}
\put(9.55,6.9){exp.}
\put(10.3,6.9){model}
\put(3,11.5){$h(t), c(t), \Delta T(t),b(t), u(t)$}
\put(4.5,6.3){model with $Fr=Fr_0$ (Eq. \ref{eq:ch3modelsolution})}
\put(4.5,1.3){model with $\frac{Fr_0^2}{Fr^2}=1+K\Lambda_sx^*_f$}
\end{picture}
       \caption{Panel (a) shows the sketch of the box model. Panels (b) and (c) compare the model predictions with the experimental results for the EXP3a inlet condition, parameterising $Fr$ as $Fr=Fr_0$ and as in equation \ref{eq:froudepar}, respectively, with $Fr_0=1.1$ and $K=7$.}
    \label{fig:boxsketch}
\end{figure}
A simple lumped analytical model is developed to predict the front position as a function of time. 
The current is modeled as a box (figure \ref{fig:boxsketch}(a)), i.e. a rectangle of length $x_f(t)$ and height $h(t)$. We assume that outside the current, the air temperature and density are homogeneous and equal to $T_0$ and $\rho_0$, respectively. Inside the current, spatially homogeneous CO$_2$ volume fraction $c(t)$ and temperature difference with respect to the ambient $\Delta T(t)=T(t)-T_0$, where $T$ is the current temperature, are considered. Within the box-model framework, we assume that all the current variables are only functions of time, and not of space. An air-CO$_2$ mixture with a CO$_2$ volume fraction $c_s$, velocity $u_s$, and temperature $T_s=T_0$ is injected from a source of height $h_s$. Since the maximum $\text{CO}_2$ volume fraction in the mixture is $0.1$, the specific heat per unit mass of the mixture can be considered equal to the air one, $c_{p}\approx c_{p,0}$, and the mass fraction in the mixture can be approximated as $\tilde c(t) \equiv (M_{\text{CO}_2}/M)c(t) \approx (M_{\text{CO}_2}/M_0)c(t)$, where $M$, $M_{\text{CO}_2}$ and $M_0$ are the mixture,  pure carbon dioxide and pure air molar masses, respectively \citep{lanzini2026}. 
Exploiting the Boussinesq approximation and the ideal gas law, the values of $c(t)$ and $\Delta T(t)$ can be combined to obtain the current buoyancy $b(t)=g(\rho(t)-\rho_0)/\rho_0$, where $\rho(t)$ is the current density, as:
\begin{equation}
    b(t)=g\bigg (\chi_M c(t) -\frac{\Delta T(t)}{T_0} \bigg),\label{eq:ch3buoyancyapprox}
\end{equation}
where $\chi_M=(M_{\text{CO}_2}-M_0)/M_0=0.519$. 
The detailed derivation of equation \ref{eq:ch3buoyancyapprox} for a heated CO$_2$-air mixture is given in \cite{lanzini2026}.

The total (per unit span) mass of carbon dioxide ${\cal M}$, excess of enthalpy with respect to the source ${\cal H}$, and buoyancy ${\cal B}$ of the current are, respectively: 
\begin{equation}
     {\cal M}(t)= \rho_0 \frac{M_{\text{CO}_2}}{M_0} c(t) h(t) x_f(t),\label{eq:ch3Mdefiniton}
\end{equation}
\begin{equation}
    {\cal H}(t)= \rho_0 c_{p,0} \Delta T(t) h(t) x_f(t), \label{eq:ch3Hdefiniton}
 \end{equation}   
\begin{equation}
    {\cal B}(t)= b(t) h(t) x_f(t) =\frac{g \chi_M}{\rho_0}\frac{M_0}{M_{\text{CO}_2}}{\cal M}(t) -\frac{g}{\rho_0 c_{p,0} T_0} {\cal H}(t),\label{eq:ch3Bdefiniton}
\end{equation}
where the right-hand side of equation (\ref{eq:ch3Bdefiniton}) has been obtained by combining equations (\ref{eq:ch3buoyancyapprox}), (\ref{eq:ch3Mdefiniton}), and (\ref{eq:ch3Hdefiniton}).
 Substituting $b(t)h(t)$ by $b_h(t)h_h(t)$ in equation (\ref{eq:ch3Bdefiniton}), i.e.,  assuming that $b$ and $h$ are equal to the head ones, and exploiting the definition of Froude number at the head (Eq. \ref{eq:Fr}),  ${\cal B}$  can also be written as:
\begin{equation}
    {\cal B}(t)=\frac{1}{Fr^2(\Lambda_s,t)}\bigg ( \frac{\text{d} x_f}{\text{d}t} \bigg)^2 x_f. \label{eq:ch3B2}
\end{equation}
The conservation of mass of carbon dioxide and enthalpy allows writing the time derivative of $\cal M$ and $\cal H$, respectively, as:

   \begin{equation}
     \frac{\text{d}\cal M}{\text {d}t}= \rho_0 \frac{M_{\text{CO}_2}}{M_0} c_s h_s u_s,\label{eq:ch3Mcons}
\end{equation}
\begin{equation}
    \frac{\text{d}\cal H}{\text {d}t}= \varphi x_f, \label{eq:ch3Hcons}
 \end{equation}  
where the right-hand side of equation (\ref{eq:ch3Mcons}) is the CO$_2$ mass flux injected at the inlet and the term $\varphi x_f$  in equation (\ref{eq:ch3Hcons}) is the power per unit span transferred to the current through the heated wall.
The total buoyancy time derivative, obtained combining equations (\ref{eq:ch3Bdefiniton}), (\ref{eq:ch3Mcons}), and (\ref{eq:ch3Hcons}) is: 
\begin{equation}
    \frac{\text{d}\cal B}{\text {d}t}= F_s \bigg( 1- \Lambda_s \frac{x_f}{h_s}\bigg), \label{eq:ch3Bcons}
 \end{equation}  
where $F_s=g \chi_M c_s u_s h_s= b_su_sh_s$ is the buoyancy flux per unit span injected at the source. 
Finally, by combining equations (\ref{eq:ch3Bcons}) and (\ref{eq:ch3B2}), a single ordinary differential equation governing the front position, $x_f$, is obtained:
\begin{equation}
    \frac{\text d }{\text d t} \bigg [ \frac{1}{Fr^2}  \bigg (\frac{\text d x_f}{\text d t} \bigg)^2 x_f\bigg ]= F_s\biggl(1-\Lambda_s \frac{x_f}{h_s} \biggr).\label{eq:ch3model}
\end{equation}
In non-dimensional form, equation (\ref{eq:ch3model}) becomes:
\begin{equation}
    \frac{\text d }{\text d t^*} \bigg [ \frac{1}{Fr^2}  \bigg (\frac{\text d x^*_f}{\text d t^*} \bigg)^2 x^*_f\bigg ]= \biggl(1-\Lambda_s x^*_f \biggr),\label{eq:ch3modeladim}
\end{equation}
where $Fr(t^*,\Lambda_s)$ is the local head Froude number. Integrating equation~(\ref{eq:ch3modeladim}) yields an implicit relation between the dimensionless front position ($x_f^*$), the local Froude number $Fr$, dimensionless time $t^*$, and $\Lambda_s$. In general, this relation does not admit a closed-form solution and must therefore be evaluated numerically, providing the desired relationship expressed in equation~(\ref{eq:goal2}).

Obtaining a reliable experimental estimate of the front Froude number through equation (\ref{eq:Fr}) is technically challenging. Such an estimate would require knowledge of the vertical buoyancy distribution within the current head, which in turn would necessitate simultaneous measurements of both $\text{CO}_2$ concentration and temperature. Moreover, in the heated cases, the buoyancy field evolves continuously as the current propagates and the head progressively warms (figure \ref{fig:visual}), implying that these measurements would need to provide both high spatial and temporal resolution. To the best of the authors' knowledge, no measurements reported in the literature provide high-frequency, spatially resolved temperature measurements in air simultaneously with a second buoyant scalar field.
Since a direct experimental determination of $Fr$ is not available, we consider two alternative parameterizations of the Froude number to solve equation (\ref{eq:ch3modeladim}). 

As a first approach, we adopt the simplest possible assumption, namely that the head Froude number remains constant throughout the propagation and is equal to the value reported by \cite{SherWoods2017} for the adiabatic case, i.e. $Fr = Fr_0 = 1.1$, independently of $Fr_s$.
The solution $x^*_f(t^*)$ obtained by integrating equation (\ref{eq:ch3modeladim}) imposing $Fr=Fr_0$ is computed for different values of $\Lambda_s$ and compared with the experimental results of case EXP3a in figure \ref{fig:boxsketch}(b). For clarity, the experimental dispersion bars are omitted in the figure. The results of this model are of particular interest for three reasons. First, assuming $Fr=Fr_0$, equation (\ref{eq:ch3modeladim}) admits the following analytical solution:
\begin{equation}
   x^*_f=\frac{5}{3\Lambda_s} \bigg \{ 1- \bigg [ 1-\frac{2}{5} Fr_0^{2/3} \Lambda_s t^* \bigg ]^{3/2} \bigg \}.\label{eq:ch3modelsolution}
\end{equation}
Second, the equation (\ref{eq:ch3modelsolution})  predicts that, consistently with the experimental results, the gravity current front eventually stops at a finite distance from the source inversely proportional to $\Lambda_s$ and independent on $Fr_0$, in particular at $x^{*\text{box}}_{\text{stop}}=5/(3\Lambda_s)$. 
The possible reasons for the discrepancy between the experimentally observed stopping position, $x_{\text{stop}}^{*}=2.3/\Lambda_s$ (figure \ref{fig:stop}(c)), are discussed at the end of this section.
Third, equation (\ref{eq:ch3modelsolution}) provides a useful framework for assessing the relative contributions of the two physical mechanisms identified as responsible for the deceleration of the gravity current, namely the warming of the current fluid itself and the increase in the wall drag associated with the interaction between the current head and the rising thermal plumes (see \S \ref{subsec:flowvisual}). Specifically, the model results $x^*_f(t^*)$ obtained from equation (\ref{eq:ch3modelsolution}) account exclusively for the direct effect of warming through the parameter $\Lambda_s$, while neglecting any increase in the wall drag. Indeed, by prescribing a constant Froude number, $Fr=Fr_0$, the additional resistance to the horizontal propagation induced by the interaction with the thermal plumes is absent from the model formulation. Consequently, equation (\ref{eq:ch3modelsolution}) predicts a faster front propagation than that observed experimentally (figure \ref{fig:boxsketch}(b)). 

As a second approach, we consider a more realistic parameterization of the Froude number $Fr$. As the gravity-current fluid progressively warms, the horizontal buoyancy flux per unit area associated with its propagation, $b(t)\,\text{d}x_f/\text{d}t$, is expected to decrease for increasing $x_f$. By contrast, the vertical buoyancy flux per unit area generated by the imposed heating, $\varphi g/(\rho_0 c_p T_0)$,
which drives the rising thermal plumes, is spatially uniform. Consequently, the resistance to the horizontal motion due to interaction with the vertical plumes is expected to increase progressively with the distance travelled by the gravity current. By analogy with adiabatic gravity currents propagating over rough surfaces, for which the head Froude number is known to decrease as the drag increases \citep{Nogueira2013,Maggi2022,maggi2025}, we assume that a similar behaviour also applies in the present configuration.
A simple parameterization that captures the progressive reduction in the Froude number, while preserving the possibility of numerically integrating equation (\ref{eq:ch3modeladim}), is:
\begin{equation}
    \frac{Fr_0^2}{Fr^2}=1+K\Lambda_sx^*_f, \label{eq:froudepar}
\end{equation}
where the reduction in the Froude number is assumed to become more pronounced for larger values of $\Lambda_s$ and $x^*_f$, while $K$ is a dimensionless free parameter of the model. 
Figure \ref{fig:boxsketch}(c) shows the numerical solution of equation (\ref{eq:ch3modeladim}), obtained by parameterizing the head Froude number through equation (\ref{eq:froudepar}) with $K=7$, the value that provides the best fit to the experimental data.
It can be shown (Appendix \ref{app:stop}) that, under this second parameterization of $Fr$, the model also predicts that the gravity-current front eventually stops its advancing motion at a position
$
x_{\text{stop}}^{*\text{box}}={\xi_{\text{stop}}}/{\Lambda_s},
$
where $\xi_{\text{stop}}$ is a numerical constant that depends only on $K$. For $K=7$, $\xi_{\text{stop}}\approx1.5$. If $x^*_f<x_{\text{stop}}^{*\text{box}}$, the agreement between the model predictions and the experimental measurements is excellent (figure \ref{fig:boxsketch}(c)). 

For both parameterizations of the Froude number considered, namely $Fr=Fr_0$ and equation (\ref{eq:froudepar}), the stopping position predicted by the model, $x_{\text{stop}}^{*\text{box}}$, is approximately $30\%$ smaller than that observed experimentally, $x_{\text{stop}}^{*}$, (see the black lines in figure \ref{fig:stop}(c)). Several factors may account for this discrepancy. The first limitation clearly arises from the box-model assumption, which greatly simplifies the actual flow dynamics. In addition, the experiment visualizations (figure \ref{fig:stop}(a,b)) show that, as the gravity current progressively slows down and approaches its stopping position, the head height increases continuously. Under these conditions, describing the flow as a gravity current becomes questionable. Furthermore, the model neglects the warming of the ambient fluid downstream of the current due to natural convection. As a consequence, the ambient temperature and density ahead of the front may differ from the reference values $T_0$ and $\rho_0$. This would increase the effective buoyancy of the injected current, which should therefore be evaluated relative to the local ambient density rather than to the reference density far from the wall $\rho_0$. The present model can also be applied to the configuration in which wall heating is restricted to the region $x<x_f$ and an adiabatic condition is present for $x>x_f$. In this case, however, the values of the head Froude number are expected to differ from those in the present configuration.

\section{Conclusion}\label{sec:conclusion}
We have experimentally investigated the propagation of sustained gravity currents along a uniformly heated horizontal wall and characterised the resulting front propagation speed as a function of the governing non-dimensional parameters.

The results show that, in addition to the source Froude number $Fr_s$, the key parameter influencing the flow dynamics is $\Lambda_s$, defined as the ratio between the vertical buoyancy flux per unit area induced by the homogeneous convective heating, $f_\varphi$, and the horizontal buoyancy flux per unit area driving the current $f_s$. As $\Lambda_s$ increases, the front velocity is reduced. This slowdown is caused by a reduction of the buoyancy flux at the head, due to the direct heating of the current fluid, and by an increased resistance to horizontal propagation resulting from the interaction between the front's horizontal motion and the thermal plumes induced by natural convection downstream of the front.  
For the conditions investigated, which are predominantly in the inertia-dominated source regime ($Fr_s>1$), the source Froude number has only a weak influence on the front position.
As the front evolution depends on the instantaneous location and intensity of the thermal plumes present in the channel at the time of current injection, the reported results are ensemble averages over repeated runs.
For the most strongly heated conditions, it is possible to stop the advance of the front. In this situation, the difference in density between the head and the air downstream is cancelled, and the head becomes a light buoyant plume. For all the source Froude numbers and for $\Lambda_s<0.1$, the experimental stop position, $x_{\text{stop}}^*$, follows the relation $x_{\text{stop}}^*=2.3/\Lambda_s$. For $\Lambda_s>0.1$, the stop position is sufficiently close to the source that it becomes sensitive to its inertia at the source, i.e., to $Fr_s$. 

The experimental results are interpreted by comparison with a lumped analytical model that predicts the front position as a function of time.  Solving the model requires the Froude number of the current head, $Fr$, to be specified. Since a direct experimental determination of $Fr$ is technically challenging, two alternative parameterizations are considered. In the first, the Froude number is assumed to remain constant, $Fr=Fr_0$, equal to the value characteristic of an adiabatic gravity current. In the second, $Fr$ is assumed to decrease during the propagation according to equation (\ref{eq:froudepar}). With both parameterizations, the model predicts that the gravity-current front eventually comes to rest at a position $x_{\mathrm{stop}}^{*\mathrm{box}}$ approximately $30\%$ smaller than that measured experimentally. This discrepancy may reflect the limited validity of the model when the front propagates very slowly, and the head height grows without bound. Under these conditions, the flow may no longer be adequately described as a gravity current. Nevertheless, for $x^*_f<x_{\mathrm{stop}}^{*\mathrm{box}}$, the model incorporating the parameterization given by equation (\ref{eq:froudepar}) is in excellent agreement with the experimental observations.

It would be of interest to investigate other non-adiabatic wall configurations, such as surface cooling, or heat exchange occurring only for positions $x < x_f$, i.e., considering an adiabatic condition downstream of the gravity-current front. The imposition of a homogeneous value of $\varphi$, which does not vary with the distance from the source, represents an ideal boundary condition that simplifies the real physical configuration.  In general, for gravity currents induced by temperature differences, the value of $\varphi$ may vary during the current propagation and may depend on the temperature difference between the current and the wall.  Future experiments and models should therefore focus on configurations in which $\varphi = \varphi(x)$. Finally, in addition to the flow visualizations presented here, it would be valuable to perform quantitative measurements of the buoyancy and velocity fields within the current head in order to obtain a direct experimental estimate of the head Froude number $Fr$ as a function of $t^*$ and $\Lambda_s$.

\backsection[Supplementary movies] {Supplementary movies of the experiments are available as supplementary material.}

\backsection[Acknowledgements]{The authors acknowledge the technical workshop of the LMFA laboratory (Service études et fabrication) and Horacio Correia for the realization of the experimental facility. S.L. thanks the undergraduate students Ganesh Tangavelou and Mohammed El-Naggar, for their contribution to the experimental campaign.}

\backsection[Funding]{This work was funded by the Région Auvergne-Rhône-Alpes (France) within the COVENTU Project (grant number: GFC
2020–043)}

\backsection[Declaration of interests] {The authors report no conflict of interest.}


\backsection[Author ORCIDs]{S. Lanzini, https://orcid.org/0000-0003-4270-8253; M. Creyssels, https://orcid.org/0000-0001-5203-3275; M. Marro, https://orcid.org/0000-0002-6661-9637; S.Vaux, https://orcid.org/0000-0002-6544-9701; P. Salizzoni  https://orcid.org/0000-0001-5987-9839}

\appendix 
\section{Model stop position if $Fr_0^2/Fr^2=1+K\Lambda_sx^*_f$}
\label{app:stop}
We show here that if the evolution of the head Froude number $Fr$ in the equation (\ref{eq:ch3modeladim}) is described by the relation $Fr_0^2/Fr^2=1+K\Lambda_sx^*_f$, then the predicted stopping position is of the form $x_{\text{stop}}^{*\text{box}}=\xi_{\text{stop}}/\Lambda_s$, where $\xi_{\text{stop}}$ is a constant that depends only on $K$.
Equation (\ref{eq:ch3modeladim}) becomes
\begin{equation}
    \frac{\text d}{\text dt^*}
    \left[
    (1+K\Lambda_s x^*_f)x^*_f
    \left(\frac{\text d x^*_f}{\text dt^*}\right)^2
    \right]
    =
    Fr_0^2(1-\Lambda_s x^*_f),
    \label{eq:model_appendix}
\end{equation}
where $\Lambda_s$, $Fr_0$ and $K$ are positive constants. Introducing $D(x^*_f)=x^*_f(1+K\Lambda_s x^*_f)$ and $z(x^*_f)=D(x^*_f)\left(\text d x^*_f/\text d t^*\right)^2$, and writing the time derivative of a generic function of $x^*_f$ as
$\text d/\text dt^*=(\text d x^*_f/\text d t^*)\,\text d/\text d x^*_f$,
equation \eqref{eq:model_appendix} becomes
\begin{equation}
    \frac{\text d x^*_f}{\text dt^*}
    \frac{\text d z}{\text d x^*_f}
    =
    Fr_0^2(1-\Lambda_s x^*_f).
    \label{eq:appendix2}
\end{equation}

Since
$
\displaystyle
\frac{\text d x^*_f}{\text dt^*}
=
\sqrt{\frac{z}{D(x^*_f)}},
$
equation \eqref{eq:appendix2} can be rewritten as
\begin{equation}
    \sqrt{z}
    \frac{\text d z}{\text d x^*_f}
    =
    Fr_0^2(1-\Lambda_s x^*_f)
    \sqrt{D(x^*_f)}.
    \label{eq:appendix3}
\end{equation}

Integrating equation \eqref{eq:appendix3}, subject to the initial condition $x^*_f(t^*=0)=0$, yields
\begin{equation}
    z(x^*_f)
    =
    \left[
    \frac{3}{2}Fr_0^2
    \int_0^{x^*_f}
    (1-\Lambda_s y)
    \sqrt{y(1+K\Lambda_s y)}
    \,\text dy
    \right]^{2/3}.
    \label{eq:appendix4}
\end{equation}

The front stops propagating when $\text d x^*_f/\text dt^*=0$, i.e. when $z(x^*_f)=0$. From equation \eqref{eq:appendix4}, this occurs when
\begin{equation}
    \int_0^{x_{\text{stop}}^{*\text{box}}}
    (1-\Lambda_s y)
    \sqrt{y(1+K\Lambda_s y)}
    \,\text dy
    =0.
    \label{eq:appendixxstop}
\end{equation}

Introducing the dimensionless variable $\xi=\Lambda_sy$, equation \eqref{eq:appendixxstop} becomes
\begin{equation}
    \int_0^{\xi_{\text{stop}}}
    (1-\xi)
    \sqrt{\xi(1+K\xi)}
    \,\text d\xi
    =0,
    \label{eq:appendixfinal}
\end{equation}
where $\xi_{\text{stop}}=\Lambda_sx_{\text{stop}}^{*\text{box}}$ is the numerical root of equation \eqref{eq:appendixfinal}. Hence, the stopping position can be written as $x_{\text{stop}}^{*\text{box}}=\xi_{\text{stop}}/\Lambda_s$, where $\xi_{\text{stop}}$ depends only on $K$. For $K=7$, $\xi_{\text{stop}}\approx1.5$.

\bibliographystyle{jfm}
\bibliography{jfm}

@article{Nogueira2013,
author = {H. I. S. Nogueira and C. Adduce and E. Alves and M. J. Franca },
title = {Analysis of lock-exchange gravity currents over smooth and rough beds},
journal = {Journal of Hydraulic Research},
volume = {51},
number = {4},
pages = {417--431},
year = {2013},
publisher = {IAHR Website},
doi = {10.1080/00221686.2013.798363},


URL = { 
    
        https://doi.org/10.1080/00221686.2013.798363
    
    

},
eprint = { 
    
        https://doi.org/10.1080/00221686.2013.798363
    
    

}

}

@Article{Maggi2022,
author={Maggi, M. R.
and Adduce, C.
and Negretti, M. E.},
title={Lock-release gravity currents propagating over roughness elements},
journal={Environmental Fluid Mechanics},
year={2022},
month={Jun},
day={01},
volume={22},
number={2},
pages={383-402},
issn={1573-1510},
doi={10.1007/s10652-022-09845-6},
url={https://doi.org/10.1007/s10652-022-09845-6}
}

@article{Pritchard1982,
author = {Pritchard, D. T. and Currie, J. A.},
title = {Diffusion of coefficients of carbon dioxide, nitrous oxide, ethylene and ethane in air and their measurement},
journal = {Journal of Soil Science},
volume = {33},
number = {2},
pages = {175-184},
doi = {https://doi.org/10.1111/j.1365-2389.1982.tb01757.x},
url = {https://bsssjournals.onlinelibrary.wiley.com/doi/abs/10.1111/j.1365-2389.1982.tb01757.x},
eprint = {https://bsssjournals.onlinelibrary.wiley.com/doi/pdf/10.1111/j.1365-2389.1982.tb01757.x},
year = {1982}
}

@article{Simpson1982,
   author = "Simpson, J E",
   title = "Gravity Currents in the Laboratory, Atmosphere, and
Ocean", 
   journal= "Annual Review of Fluid Mechanics",
   year = "1982",
   volume = "14",
   number = "Volume 14, 1982",
   pages = "213-234",
   doi = "https://doi.org/10.1146/annurev.fl.14.010182.001241",
   url = "https://www.annualreviews.org/content/journals/10.1146/annurev.fl.14.010182.001241",
   publisher = "Annual Reviews",
   issn = "1545-4479",
   type = "Journal Article",
  }

@article{Simpson1977,
author = {Simpson, J. E. and Mansfield, D. A. and Milford, J. R.},
title = {Inland penetration of sea-breeze fronts},
journal = {Quarterly Journal of the Royal Meteorological Society},
volume = {103},
number = {435},
pages = {47-76},
doi = {https://doi.org/10.1002/qj.49710343504},
url = {https://rmets.onlinelibrary.wiley.com/doi/abs/10.1002/qj.49710343504},
eprint = {https://rmets.onlinelibrary.wiley.com/doi/pdf/10.1002/qj.49710343504},
year = {1977}
}

@Article{Jiang2018,
author={Jiang, Y.
and Liu, X.},
title={Experimental and numerical investigation of density current over macro-roughness},
journal={Environmental Fluid Mechanics},
year={2018},
month={Feb},
day={01},
volume={18},
number={1},
pages={97-116},
issn={1573-1510},
doi={10.1007/s10652-016-9500-1},
url={https://doi.org/10.1007/s10652-016-9500-1}
}

@article{Horsley2018, title={A note on analytic solutions for entraining stratified gravity currents}, volume={836}, DOI={10.1017/jfm.2017.834}, journal={Journal of Fluid Mechanics}, author={Horsley, M. C. and Woods, A. W.}, year={2018}, pages={260–276}}

@article { CenedeseMonti2003,
      author = "A. Cenedese and P. Monti",
      title = "Interaction between an Inland Urban Heat Island and a Sea-Breeze Flow: A Laboratory Study",
      journal = "Journal of Applied Meteorology",
      year = "2003",
      publisher = "American Meteorological Society",
      address = "Boston MA, USA",
      volume = "42",
      number = "11",
      doi = "10.1175/1520-0450(2003)042<1569:IBAIUH>2.0.CO;2",
      pages=      "1569 - 1583",
      url = "https://journals.ametsoc.org/view/journals/apme/42/11/1520-0450_2003_042_1569_ibaiuh_2.0.co_2.xml"
}

@article{Linden1999,
   author = {Linden, P. F.},
   title = {THE FLUID MECHANICS OF NATURAL VENTILATION}, 
   journal= {Annu. Rev. Fluid Mech.},
   year = {1999},
   volume = {31},
   number = {Volume 31, 1999},
   pages = {201-238},
   doi = {https://doi.org/10.1146/annurev.fluid.31.1.201},
   url = {https://www.annualreviews.org/content/journals/10.1146/annurev.fluid.31.1.201}
}

@article{MAXWORTHY2002, title={The propagation of a gravity current into a linearly stratified fluid}, volume={453}, DOI={10.1017/S0022112001007054}, journal={Journal of Fluid Mechanics}, author={Maxworthy, T. and Leilich, J. and Simpson, J. E. and Meiburg, E. H.}, year={2002}, pages={371–394}}

@article{Longo2016,
    author = {Longo, S. and Ungarish, M. and Di Federico, V. and Chiapponi, L. and Addona, F.},
    title = {Gravity currents in a linearly stratified ambient fluid created by lock release and influx in semi-circular and rectangular channels},
    journal = {Physics of Fluids},
    volume = {28},
    number = {9},
    pages = {096602},
    year = {2016},
    month = {09},
    issn = {1070-6631},
    doi = {10.1063/1.4963009},
    url = {https://doi.org/10.1063/1.4963009},
    eprint = {https://pubs.aip.org/aip/pof/article-pdf/doi/10.1063/1.4963009/14835585/096602\_1\_online.pdf},
}

@article{Chiapponi2018,
title={Critical regime of gravity currents flowing in non-rectangular channels with density stratification}, volume={840}, DOI={10.1017/jfm.2017.917}, journal={Journal of Fluid Mechanics}, author={Chiapponi, L. and Ungarish, M. and Longo, S. and Di Federico, V. and Addona, F.}, year={2018}, pages={579–612}}

@article{Ungarish2023,
  title = {Strongly supercritical non-Boussinesq sustained gravity currents: Time-dependent and steady-state approximate solutions},
  author = {Ungarish, M.},
  journal = {Phys. Rev. Fluids},
  volume = {8},
  issue = {5},
  pages = {053801},
  numpages = {21},
  year = {2023},
  month = {May},
  publisher = {American Physical Society},
  doi = {10.1103/PhysRevFluids.8.053801},
  url = {https://link.aps.org/doi/10.1103/PhysRevFluids.8.053801}
}

@article{CenedeseAdduce2008, title={Mixing in a density-driven current flowing down a slope in a rotating fluid}, volume={604}, DOI={10.1017/S0022112008001237}, journal={Journal of Fluid Mechanics}, author={Cenedese, C. and Adduce, C.}, year={2008}, pages={369–388}}

@article{LindenSimpson1986, title={Gravity-driven flows in a turbulent fluid}, volume={172}, DOI={10.1017/S0022112086001829}, journal={Journal of Fluid Mechanics}, author={Linden, P. F. and Simpson, J. E.}, year={1986}, pages={481–497}}

@article{Thompsonetal,
author = {Thompson, W. T. and Holt, T. and Pullen, J.},
title = {Investigation of a sea breeze front in an urban environment},
journal = {Quarterly Journal of the Royal Meteorological Society},
volume = {133},
number = {624},
pages = {579-594},
doi = {https://doi.org/10.1002/qj.52},
url = {https://rmets.onlinelibrary.wiley.com/doi/abs/10.1002/qj.52},
eprint = {https://rmets.onlinelibrary.wiley.com/doi/pdf/10.1002/qj.52},
year = {2007}
}

@article{Yoshikado1994,
  title={Interaction of the Sea Breeze with Urban Heat Islands of Different Sizes and Locations},
  author={Yoshikado, H.},
  journal={Journal of the Meteorological Society of Japan. Ser. II},
  volume={72},
  number={1},
  pages={139-143},
  year={1994},
  doi={10.2151/jmsj1965.72.1_139}
}

@article { Yoshikado1992,
      author = "Yoshikado,H.",
      title = "Numerical Study of the Daytime Urban Effect and Its Interaction with the Sea Breeze",
      journal = "Journal of Applied Meteorology and Climatology",
      year = "1992",
      publisher = "American Meteorological Society",
      address = "Boston MA, USA",
      volume = "31",
      number = "10",
      doi = "10.1175/1520-0450(1992)031<1146:NSOTDU>2.0.CO;2",
      pages=      "1146 - 1164",
      url = "https://journals.ametsoc.org/view/journals/apme/31/10/1520-0450_1992_031_1146_nsotdu_2_0_co_2.xml"
}

@book{Barenblatt_1996,
place={Cambridge}, 
series={Cambridge Texts in Applied Mathematics}, title={Scaling, Self-similarity, and Intermediate Asymptotics: Dimensional Analysis and Intermediate Asymptotics}, publisher={Cambridge University Press}, 
number={1},
author={Barenblatt, G. I.}, year={1996},
collection={Cambridge Texts in Applied Mathematics}}

@article{Massman1998,
title = {A review of the molecular diffusivities of H2O, CO2, CH4, CO, O3, SO2, NH3, N2O, NO, and NO2 in air, O2 and N2 near STP},
journal = {Atmospheric Environment},
volume = {32},
number = {6},
pages = {1111-1127},
year = {1998},
issn = {1352-2310},
doi = {https://doi.org/10.1016/S1352-2310(97)00391-9},
url = {https://www.sciencedirect.com/science/article/pii/S1352231097003919},
author = {W. J. Massman}
}

@article{Turner1974,
   author = "Turner, J. S.",
   title = "Double-Diffusive Phenomena", 
   journal= "Annual Review of Fluid Mechanics",
   year = "1974",
   volume = "6",
   number = "Volume 6, 1974",
   pages = "37-54",
   doi = "https://doi.org/10.1146/annurev.fl.06.010174.000345",
   url = "https://www.annualreviews.org/content/journals/10.1146/annurev.fl.06.010174.000345",
   publisher = "Annual Reviews",
   issn = "1545-4479",
   type = "Journal Article",
  }

@article{SherWoods2017, title={Mixing in continuous gravity currents}, volume={818}, DOI={10.1017/jfm.2017.168}, journal={Journal of Fluid Mechanics}, author={Sher, D. and Woods, A. W.}, year={2017}, pages={R4}}

@article{Hoggetal2005, title={On gravity currents driven by constant fluxes of saline and particle-laden fluid in the presence of a uniform flow}, volume={539}, DOI={10.1017/S002211200500546X}, journal={Journal of Fluid Mechanics}, author={Hogg, A. J. and Hallworth, M. A. and Huppert, H. E.}, year={2005}, pages={349–385}}

@article{Salizzoni2018,
title = {Influence of source conditions and heat losses on the upwind back-layering flow in a longitudinally ventilated tunnel},
journal = {International Journal of Heat and Mass Transfer},
volume = {117},
pages = {143-153},
year = {2018},
issn = {0017-9310},
doi = {https://doi.org/10.1016/j.ijheatmasstransfer.2017.10.017},
url = {https://www.sciencedirect.com/science/article/pii/S0017931017304982},
author = {P. Salizzoni and M. Creyssels and L. Jiang and A. Mos and R. Mehaddi and O. Vauquelin}
}

@article{Hoggetal2016, title={Sustained gravity currents in a channel}, volume={798}, DOI={10.1017/jfm.2016.343}, journal={Journal of Fluid Mechanics}, author={Hogg, A. J. and Nasr-Azadani, M. M. and Ungarish, M. and Meiburg, E.}, year={2016}, pages={853–888}}

@article{SherWoods2015, title={Gravity currents: entrainment, stratification and self-similarity}, volume={784}, DOI={10.1017/jfm.2015.576}, journal={Journal of Fluid Mechanics}, author={Sher, D. and Woods, A. W.}, year={2015}, pages={130–162}}

@article{Nogueira_2013,
doi = {10.1088/0957-0233/24/4/047001},
url = {https://dx.doi.org/10.1088/0957-0233/24/4/047001},
year = {2013},
month = {mar},
publisher = {IOP Publishing},
volume = {24},
number = {4},
pages = {047001},
author = {Nogueira, H. I. S. and Adduce, C. and Alves, E. and Franca, M. J.},
title = {Image analysis technique applied to lock-exchange gravity currents},
journal = {Measurement Science and Technology},

}

@article{Townsend_1972, title={Mixed convection over a heated horizontal plane}, volume={55}, DOI={10.1017/S0022112072001818}, number={2}, journal={Journal of Fluid Mechanics}, author={Townsend, A. A.}, year={1972}, pages={209–227}}

@article{Simpson_Britter_1979, title={The dynamics of the head of a gravity current advancing over a horizontal surface}, volume={94}, DOI={10.1017/S0022112079001142}, number={3}, journal={Journal of Fluid Mechanics}, author={Simpson, J. E. and Britter, R. E.}, year={1979}, pages={477–495}}

@article{Griffiths1986,
   author = "Griffiths, R. W.",
   title = "Gravity Currents in Rotating Systems", 
   journal= "Annual Review of Fluid Mechanics",
   year = "1986",
   volume = "18",
   number = "Volume 18, 1986",
   pages = "59-89",
   doi = "https://doi.org/10.1146/annurev.fl.18.010186.000423",
   url = "https://www.annualreviews.org/content/journals/10.1146/annurev.fl.18.010186.000423",
   publisher = "Annual Reviews",
   issn = "1545-4479",
   type = "Journal Article",
  }

@article{GriffithsHopfinger1983, title={Gravity currents moving along a lateral boundary in a rotating fluid}, volume={134}, DOI={10.1017/S0022112083003407}, journal={Journal of Fluid Mechanics}, author={Griffiths, R. W. and Hopfinger, E. J.}, year={1983}, pages={357–399}}

@article { Cenedese2004,
      author = "C. Cenedese and J. A. Whitehead and T. A. Ascarelli and M. Ohiwa",
      title = "A Dense Current Flowing down a Sloping Bottom in a Rotating Fluid",
      journal = "Journal of Physical Oceanography",
      year = "2004",
      publisher = "American Meteorological Society",
      address = "Boston MA, USA",
      volume = "34",
      number = "1",
      doi = "10.1175/1520-0485(2004)034<0188:ADCFDA>2.0.CO;2",
      pages=      "188 - 203",
      url = "https://journals.ametsoc.org/view/journals/phoc/34/1/1520-0485_2004_034_0188_adcfda_2.0.co_2.xml"
}

@Article{Vidali2025,
author={Vidali, C.
and Marro, M.
and Gostiaux, L.
and Houssin, D.
and Vyazmina, E.
and Salizzoni, P.},
title={Wind-Tunnel Experiment of Heavy Gas and Passive Scalar Emission in a Turbulent Boundary Layer},
journal={Boundary-Layer Meteorology},
year={2025},
month={Mar},
day={22},
volume={191},
number={4},
pages={18},
issn={1573-1472},
doi={10.1007/s10546-025-00909-w},
url={https://doi.org/10.1007/s10546-025-00909-w}
}

@book{Simpson1997,
  author    = {J. E. Simpson},
  title     = {Gravity Currents in the Environment and the Laboratory},
  edition   = {2nd},
  year      = {1997},
  publisher = {Cambridge University Press},
  address   = {Cambridge}
}

@article{Britter_Simpson_1978, title={Experiments on the dynamics of a gravity current head}, volume={88}, DOI={10.1017/S0022112078002074}, number={2}, journal={Journal of Fluid Mechanics}, author={Britter, R. E. and Simpson, J. E.}, year={1978}, pages={223–240}}

@article{Britter_Linden_1980, title={The motion of the front of a gravity current travelling down an incline}, volume={99}, DOI={10.1017/S0022112080000754}, number={3}, journal={Journal of Fluid Mechanics}, author={Britter, R. E. and Linden, P. F.}, year={1980}, pages={531–543}}

@article{Britter1989,
   author = "Britter, R E",
   title = "Atmospheric Dispersion of Dense Gases", 
   journal= "Annual Review of Fluid Mechanics",
   year = "1989",
   volume = "21",
   number = "Volume 21, 1989",
   pages = "317-344",
   doi = "https://doi.org/10.1146/annurev.fl.21.010189.001533",
   url = "https://www.annualreviews.org/content/journals/10.1146/annurev.fl.21.010189.001533",
   publisher = "Annual Reviews",
   issn = "1545-4479",
   type = "Journal Article",
  }

@article{maggi2025,
  title={Dynamics and mixing of gravity currents over an array of cylindrical obstacles},
  author={Maggi, M. R. and Di Lollo, G. and Adduce, C.},
  journal={Physics of Fluids},
  volume={37},
  number={7},
  year={2025},
  publisher={AIP Publishing}
}

@article{laForgia,
    author = {la Forgia, G. and Ottolenghi, L. and Adduce, C. and Falcini, F.},
    title = {Intrusions and solitons: Propagation and collision dynamics},
    journal = {Physics of Fluids},
    volume = {32},
    number = {7},
    pages = {076605},
    year = {2020},
    month = {07},
    issn = {1070-6631},
    doi = {10.1063/5.0011604},
    url = {https://doi.org/10.1063/5.0011604},

}

@article{Martin2020,
  title = {Propagation of a continuously supplied gravity current head down bottom slopes},
  author = {Martin, A. and Negretti, M. E. and Ungarish, M. and Zemach, T.},
  journal = {Phys. Rev. Fluids},
  volume = {5},
  issue = {5},
  pages = {054801},
  numpages = {12},
  year = {2020},
  month = {May},
  publisher = {American Physical Society},
  doi = {10.1103/PhysRevFluids.5.054801},
  url = {https://link.aps.org/doi/10.1103/PhysRevFluids.5.054801}
}

@article{Negretti2017, title={Development of gravity currents on rapidly changing slopes}, volume={833}, DOI={10.1017/jfm.2017.696}, journal={Journal of Fluid Mechanics}, author={Negretti, M. E. and Flòr, J.-B. and Hopfinger, E. J.}, year={2017}, pages={70–97}}

@article{wilson2017,
    author = {Wilson, R. I. and Friedrich, H. and Stevens, C.},
    title = {Turbulent entrainment in sediment-laden flows interacting with an obstacle},
    journal = {Physics of Fluids},
    volume = {29},
    number = {3},
    pages = {036603},
    year = {2017},
    month = {03},
    issn = {1070-6631},
    doi = {10.1063/1.4979067},
    url = {https://doi.org/10.1063/1.4979067},
    }

@article{DEFALCO2021,
title = {On the dynamics of quasi-steady gravity currents flowing up a slope},
journal = {Advances in Water Resources},
volume = {147},
pages = {103791},
year = {2021},
issn = {0309-1708},
doi = {https://doi.org/10.1016/j.advwatres.2020.103791},
url = {https://www.sciencedirect.com/science/article/pii/S0309170820305042},
author = {M.C. {De Falco} and C. Adduce and M.E. Negretti and E.J. Hopfinger}
}

@article{Lanzini2026,
  author    = {Lanzini, S. and Marro, M. and Creyssels, M. and Azouzi, A. and Salizzoni, P.},
  title     = {Experimental study on gravity currents flowing on heated walls},
  journal   = {Experiments in Fluids},
  year      = {2026},
  volume    = {67},
  number    = {7},
  pages     = {90},
  doi       = {10.1007/s00348-026-04238-7},
  url       = {https://doi.org/10.1007/s00348-026-04238-7},
  issn      = {1432-1114}
}

@article{Harrouk_Mehaddi_Arcen_Dossmann_2025, title={The energetics of mixing in continuous gravity currents}, volume={1021}, DOI={10.1017/jfm.2025.10759}, journal={Journal of Fluid Mechanics}, author={Harrouk, M. and Mehaddi, R. and Arcen, B. and Dossmann, Y.}, year={2025}, pages={R2}}

@article{MOFFAT1988,
title = {Describing the uncertainties in experimental results},
journal = {Experimental Thermal and Fluid Science},
volume = {1},
number = {1},
pages = {3-17},
year = {1988},
issn = {0894-1777},
doi = {https://doi.org/10.1016/0894-1777(88)90043-X},
url = {https://www.sciencedirect.com/science/article/pii/089417778890043X},
author = {Robert J. Moffat}
}

@article{FOX2022,
title = {Overview of the Jack Rabbit II (JR II) field experiments and summary of the methods used in the dispersion model comparisons},
journal = {Atmospheric Environment},
volume = {269},
pages = {118783},
year = {2022},
issn = {1352-2310},
doi = {https://doi.org/10.1016/j.atmosenv.2021.118783},
url = {https://www.sciencedirect.com/science/article/pii/S1352231021006051},
author = {S. Fox and S. Hanna and T. Mazzola and T. Spicer and J. Chang and S. Gant}
}

@article{HANNA2012,
title = {The Jack Rabbit chlorine release experiments: Implications of dense gas removal from a depression and downwind concentrations},
journal = {Journal of Hazardous Materials},
volume = {213-214},
pages = {406-412},
year = {2012},
issn = {0304-3894},
doi = {https://doi.org/10.1016/j.jhazmat.2012.02.013},
url = {https://www.sciencedirect.com/science/article/pii/S0304389412001598},
author = {S. Hanna and R. Britter and E. Argenta and J. Chang}
}

@book{ungarish2020,
author = {Ungarish, M.},
title = {Gravity Currents and Intrusions},
publisher = {WORLD SCIENTIFIC},
year = {2020},
doi = {10.1142/11986},
address = {},
edition   = {},
URL = {https://www.worldscientific.com/doi/abs/10.1142/11986},
eprint = {https://www.worldscientific.com/doi/pdf/10.1142/11986}
}

@article{DiBernardino2022,
  author = {Di Bernardino, A. and Mazzarella, V. and Pecci, M. and Casasanta, G. and Cacciani, M. and Ferretti, R.},
  title = {Interaction of the Sea Breeze with the Urban Area of Rome: WRF Mesoscale and WRF Large-Eddy Simulations Compared to Ground-Based Observations},
  journal = {Boundary-Layer Meteorology},
  year = {2022},
  volume = {185},
  number = {3},
  pages = {333--363},
  doi = {10.1007/s10546-022-00734-5},
  url = {https://doi.org/10.1007/s10546-022-00734-5},
  issn = {1573-1472}
}

@article { Chen2025,
      author = "G. Chen and X. Kong and T. Takemi and K. Saito and H. Seko and J. Ito and T. Iwasaki",
      title = "Mesoscale-to-LES Modeling of the Sea-Breeze Front and Its Interaction with Turbulent Flows over a Coastal City",
      journal = "Monthly Weather Review",
      year = "2025",
      publisher = "American Meteorological Society",
      address = "Boston MA, USA",
      volume = "153",
      number = "10",
      doi = "10.1175/MWR-D-24-0276.1",
      pages=      "1909 - 1923",
      url = "https://journals.ametsoc.org/view/journals/mwre/153/10/MWR-D-24-0276.1.xml"
}
\end{document}